\documentclass[aps,prb,reprint,superscriptaddress]{revtex4-1}
\usepackage{braket}
\usepackage{graphicx}
\usepackage{amsmath}
\usepackage{amssymb}
\usepackage{comment}
\usepackage{bbold}
\usepackage[usenames]{color}

\renewcommand{\vec}[1]{\mathbf{#1}}
\usepackage{bm}

\begin{document}

\title{Spin-charge conversion in disordered two-dimensional electron gases \\
 lacking inversion symmetry}

\author{Chunli Huang}
\affiliation{
Division of Physics and Applied Physics, School of Physical and Mathematical Sciences,
Nanyang Technological University, Singapore 637371, Singapore}
\affiliation{Department of Physics, National Tsing Hua University, Hsinchu 30013, Taiwan}

\author{Mirco Milletar\`i}
\affiliation{Dipartimento di Matematica e Fisica, Università Roma Tre, 00146 Rome,
Italy }
\affiliation{Bioinformatics Institute, Agency for Science, Technology and Research
(A{*}STAR), Singapore 138671, Singapore}

\author{Miguel A. Cazalilla}
\affiliation{Department of Physics, National Tsing Hua University and National Center for Theoretical Sciences (NCTS), Hsinchu 30013, Taiwan}
\affiliation{Donostia International Physics Center (DIPC), Manuel de Lardizabal 4, 20018 San Sebastian, Spain}

\date{\today}

\begin{abstract}
We study the spin-charge conversion mechanisms in a two-dimensional gas of electrons moving in a smooth  disorder potential by accounting for both Rashba-type  and Mott's skew scattering contributions. We find that the quantum interference effects between spin-flip and skew scattering  give rise to anisotropic spin precession scattering (ASP), a direct spin-charge conversion mechanism that was  discovered in  an earlier study of graphene decorated with adatoms [C. Huang \emph{et al.} Phys.~Rev.~B \textbf{94} 085414.~(2016)]. Our findings suggest that, together with other spin-charge conversion mechanisms such as the inverse galvanic effect, ASP is a fairly universal phenomenon that should be present in disordered two-dimensional systems lacking inversion symmetry.  
\end{abstract}

\maketitle

\section{Introduction}
The possibility of  designing  spintronic devices that are entirely electrically controlled~\cite{soumyanarayanan2016emergent,wunderlich2010spin} 
is attracting much attention due to the intriguing phenomena relating spin and charge transport in spin-orbit coupled (SOC) two dimensional (2D) materials and interfaces.~\cite{sinova2015spin,rashba1984,Edelstein1990233}
Among these phenomena, the spin-Hall effect (SHE)~\cite{sinova2004universal,hirsch1999spin,sinova2015spin} is likely the best known example due to its close relationship to the anomalous Hall effect. In most systems exhibiting the SHE, the spin Hall conductivity receives both intrinsic and extrinsic contributions. 
The intrinsic contribution arises from SOC potentials respecting the translation symmetry of the crystal lattice and therefore modify  the band-structure of the material.  On the other hand,  the extrinsic contribution originates from impurities and other kinds of disorder SOC potentials that break  lattice translation symmetry and lead to momentum and spin relaxation~\cite{Nagaosa_RevModPhys.82.1539}. 

\begin{figure}[t] 
\includegraphics[width=0.45\textwidth]{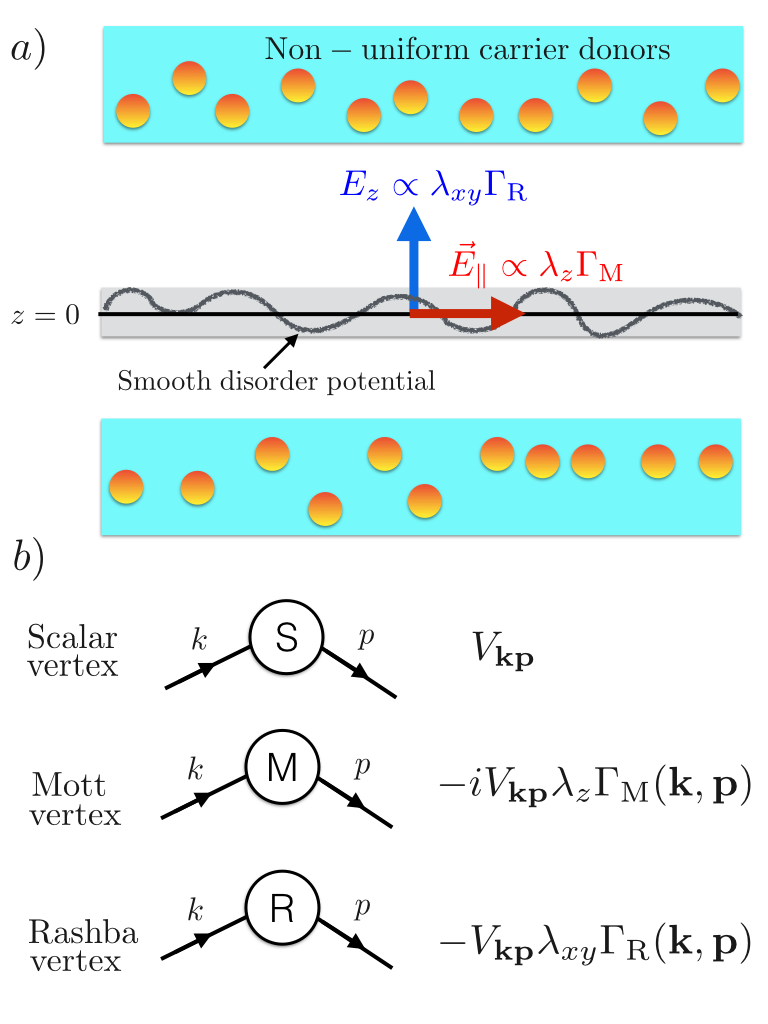}
\caption{a) The dopants in the quantum well create a smooth disorder landscape upon which the carriers in the two-dimensional electron gas move. The  perpendicular (parallel)  component of the electric field gives rise to a SOC potential of the Rashba (Mott) type. b) The Feynman diagrams for the self-energy are obtained from three types of scattering vertices describing the scattering events with the spin orbit coupling disorder potential. The vertices are the spin-independent scalar vertex, the Mott scattering vertex $\Gamma_{\mathrm{M}}(\vec{\hat{k}},\vec{\hat{p}})=  (\vec{\hat{k}}\times\vec{\hat{p}})\cdot\boldsymbol{\sigma}$ which conserves spin in the $z$-direction and the Rashba SOC vertex  $ \Gamma_{\mathrm{R}}(\vec{\hat{p}})=( \boldsymbol{\sigma} \times \vec{\hat{p}}) \cdot \hat{\vec{z}}$ which does not conserve in the $z$-direction.
}\label{fig:schematic}
\end{figure}

In addition to the SHE, systems lacking spatial inversion symmetry exhibit a closely related phenomenon, namely the current-induced spin polarization (CISP), also known as the inverse spin galvanic effect~\cite{fert_zhang_2016,amin2016the,song2016spin,song2017observation,cosimo2017theory,PhysRevLett.93.226602,ganichev2016spin}.
Similar to the SHE, CISP can arise from  both intrinsic and extrinsic mechanisms.  The intrinsic mechanism was studied many systems such as semiconductor quantum wells  with uniform Rashba SOC \cite{Edelstein1990233} and graphene proximity to transition metal dichalcogenide \cite{manuel2017}. For a 2D electron gas such as doped graphene decorated with adatoms~\cite{netoguinea09,weeks2011engineering, ferreira2014extrinsic,chunli2016direct, mirco2016crossover,mirco2016quantum}, the extrinsic mechanisms were described in a recent study by some of the present authors~\cite{chunli2016direct}. There,  a novel scattering mechanism termed  anisotropic spin precession scattering (ASP) was reported and found to yield a sizable contribution to the CISP~\cite{chunli2016direct}. More recently,  ASP  has also been shown to give rise to an anomalous nonlocal resistance in Hall bar devices~\cite{chunli2017anomalous}. ASP is a form of \emph{direct} magneto-electric effect, which arises as a quantum interference effect when an electron scatters off a single impurity that locally induces SOC by proximity.   Physically, it corresponds to a polarization of the electron spin caused by scattering with the impurities. 
 The existence of ASP scattering requires the impurity potential to break spatial inversion, which in a 2D system means that the fluctuating electric field giving rise to the SOC potential has components both in and out of the plane of the system (cf. Fig.\ref{fig:schematic}a).  Indeed,  this condition is expected to be fulfilled  in most disordered systems lacking spatial inversion symmetry. Yet, for reasons unknown to us, disorder effects that break inversion symmetry have been largely ignored when discussing the spin-charge conversion mechanisms.  Note that the effect of Rashba SOC and disorder has also been discussed intensively in the broader context of spintronics in superconductors, see Ref.\onlinecite{PhysRevB.94.180502,tokatly2017usadel,PhysRevB.93.214502,PhysRevB.94.134506}.

In graphene decorated with adatoms,~\cite{balakrishnan_colossal,chunli2016direct}  charge carriers can undergo resonant scattering with  localized impurities,~\cite{chunli2016direct} which enhances the SHE even in the dilute impurity limit~\cite{ferreira2014extrinsic, mirco2016crossover,mirco2016quantum}. However,  in many 2D systems  disorder  is not well described by a superposition of well-localized impurity potentials.  A well-known example is 2D electron gas (2DEG) in a semiconducting quantum well. In 2DEG, electrons experience a smooth disorder potential landscape arising from distant dopant impurities~\cite{glazov2010two}, for which  defining impurity density may be difficult.  Another example includes heterostructures made by placing doped graphene on a substrate such as a transition metal dichalcogenide~\cite{wang2016origin,Yang_G_on_TMD2016}. 
Even if the substrate is brought into  close contact with  graphene (in order to maximize the proximity-induced SOC), due to ripples~\cite{neto2009electronic,das2011electronic}, crystal
lattice mismatch as well as misalignments,  substrate defects and impurities,  the resulting SOC potential -- albeit smooth --  is expected to exhibit spatial fluctuations and lack spatial inversion symmetry. 

 In the case of 2DEG in quantum wells, the disorder potential results from an inhomogeneous distribution of dopant ions in the doping layer, which are typically located at a distance much larger than the atomic scale ($\sim 10$nm of separation,~\cite{tsui1982FQH} see Fig.~\ref{fig:schematic}a). The spatial gradient of the smooth disorder potential leads to a fluctuating electric field, which in turn gives rise to a disorder SOC potential.~\cite{glazov2010two} The electric field has two components, one \textit{parallel} to the plane of the 2DEG, which gives rise to a SOC that conserves the spin-projection on the axis perpendicular to the plane  and leads to Mott skew scattering. The other component of the electric field is \textit{perpendicular} to the plane of the 2DEG and will generically break the spatial inversion symmetry  (i.e. $z\rightarrow -z$). This component of the electric field gives rise to spin-flip scattering mediated by Rashba SOC~\cite{sherman2002,glazov2010two,tarasenko2006spin,tarasenko2006scattering}. Previous theoretical treatments of spin transport in the 2DEG have focused on either the Rashba-type contribution to the SOC potential~\cite{shen2014theory,raimondi2012su2, amin2016the,Nagaosa_RevModPhys.82.1539,sinova2015spin}  or  the Mott's skew-scattering~\cite{sherman2002,glazov2010two,tarasenko2006spin,tarasenko2006scattering}.
However, in this article, we shall develop a theory of spin transport in the 2DEG that treats both  disorder-induced spin-flip scattering and skew scattering on equal footing. As we argue below, this is indeed necessary in order to provide a comprehensive description of the extrinsic spin-charge conversion mechanisms.  We identify various spin-charge conversion mechanisms due to quantum mechanical interference between the different contributions of the SOC disorder potential. In particular, quantum interference between the perpendicular and the parallel component of the disorder potential is shown to give rise to ASP scattering~\cite{chunli2016direct}. Our findings suggest  that such disorder induced, quantum interference effects should be ubiquitous in any  disordered  2D material lacking spatial inversion symmetry. \\
The rest of the article is organized as follows: In Sec.~\ref{sec:setup}, we introduce the microscopic model and discuss its relationship with previously studied models. In Sec.~\ref{sec:QBE}, we discuss the  derivation of the quantum Boltzmann equation within the SU$(2)$ Schwinger-Keldysh formalism introduced in Refs.~\onlinecite{raimondi2012su2,shen2014theory,PhysRevB.74.035340} and the approximations used to obtain the collision integrals. The resulting linear response relationships are discussed in Sec.~\ref{sec:result1}. The physical consequences of our findings are discussed in Sec.~\ref{sec:result2}, with emphasis on the current-induced spin-polarization and the spin Hall effect. Then in Sec.~\ref{sec:exp}, we provide some estimation of our main results using the experiment data from Ref.~\onlinecite{bindel2016probing}. Finally, we close the article with a summary. Technical details concerning the derivation of the quantum Boltzmann equation are presented in the appendix.
\section{Microscopic Model} \label{sec:setup}
 The  Hamiltonian of the model studied below can be  written as 
\begin{align}
H=&\frac{\left( \vec{p}  + \sum_{a} \boldsymbol{\mathcal{A}}^a  \sigma^a /2 \right)^2}{2m} +  U(\vec{r}),
\end{align}
where $\vec{p}= (p_x, p_y)$ (momentum) and $\vec{r}=(x,y)$ (position) are two dimensional vectors lying in the XY plane  (i.e. $z  = 0$) to which the 2DEG is confined.
The Pauli matrices $\sigma^a$ ($a=x,y,z$) describe the electron spin. $\boldsymbol{\mathcal{A}}$ is the non-abelian gauge field describing the uniform SOC; in our model, its non-vanishing components are $\mathcal{A}_{x}^{y}=-\mathcal{A}_{y}^x=-2m\alpha$ where $\alpha$ is the potential  strength of the uniform Rashba SOC. 
The upper  and lower indices refer to the spin and orbital degree of freedom, respectively. In this article, we shall use units where $\hbar=e=1$.

 We assume the electrons move in a random disorder potential, $U(\vec{r})$, which is generated by e.g. the dopants in the quantum well~\cite{glazov2010two}. Generically,  besides a spin-independent potential, $U(\vec{r})$ contains a SOC  potential, which consists of a term accounting for Mott scattering~\cite{shen2014theory,Nagaosa_RevModPhys.82.1539,sinova2015spin} and a Rashba-type potential~\cite{sherman2002,glazov2010two,tarasenko2006spin,tarasenko2006scattering}. Mathematically,
\begin{align}\label{eq:U_dis}
U&(\vec{r})= V(\vec{r},0)-\frac{\lambda_{\parallel}^2}{4} (\vec{p} \times \boldsymbol{\nabla}_{} V(\vec{r},0))\cdot \boldsymbol{\sigma } \nonumber \\
+&\frac{\lambda_{\perp}^2}{8} \left(  \sigma_{x} \{ p_y\, ,\, \partial_z V(\vec{r},0) \} -\sigma_{y} \{ p_x\, ,\, \partial_z V(\vec{r},0) \}\right). 
\end{align}
Here $V(\vec{r},z)$  is the  fluctuating  electric potential created (in the three dimensional region of the material that contains the 2DEG) by the dopant impurities;   $\lambda_{\parallel}$ and $\lambda_{\perp}$ are the (material dependent) effective Compton-wavelengths. The second (third) term on the right hand side of Eq.~\eqref{eq:U_dis}  stems for the component of the electric field parallel (perpendicular) to the XY plane and  corresponds to the Mott (Rashba)  potential.  
The breaking of spatial inversion symmetry from disorder effects are described by the Rashba disorder potential (i.e.~last term in Eq.~\eqref{eq:U_dis}). The effect of this term on the carrier spin relaxation  was studied in~Refs.~\onlinecite{sherman2002,glazov2010two,tarasenko2006spin,tarasenko2006scattering}.  For a smooth disorder
potential, we can take $\partial_z V(\vec{r},0)\simeq V(\vec{r},0)/\xi + g(\vec{r})$, where $\xi$ is the transverse confinement length of the 2DEG. Here $g(\vec{r})$ represents the component of the perpendicular electric field that is uncorrelated to the in the plane potential, i.e.~$\langle g(\vec{r})V(\vec{r},0)\rangle=0$, see e.g. the supplementary Materials of Ref.~\onlinecite{bindel2016probing}. Note that $g(\vec{r})$ shifts the Elliott-Yafet relaxation time but it does not induce any spin-charge coupling since $\langle g(\vec{r})V(\vec{r},0)\rangle=0$.  Therefore, to the leading order in $ V(\vec{r},0)/( g(\vec{r})\xi )\gg 1$, the matrix elements of the disorder potential in the plane-wave basis can be evaluated to yield 
\begin{equation}
\langle \vec{k}| U|\vec{p} \rangle = V_{\vec{k} \vec{p}} \left[ 1- i \, \lambda_{z} \, \Gamma_{\mathrm{M}} \, (\vec{\hat{k}},\vec{\hat{p}}) - \lambda_{xy} \, \Gamma_{\mathrm{R}}(\vec{\hat{k}}+\vec{\hat{p}})  \right],
\end{equation}
where $\vec{\hat{k}} = \vec{k}/k$, and $\vec{\hat{p}} = \vec{p}/p$ and
\begin{equation}
V_{\vec{kp}} = V_{\vec{k}-\vec{p}}=\frac{1}{A}\int d\boldsymbol{r} \, e^{i (\vec{p}-\vec{k})\cdot \vec{r}} \, V(\vec{r},0)
\end{equation}
is  the Fourier transform of the in-plane disorder electric potential. 
From here and what follows, we set the area of the 2D electron gas $A=1$. $\Gamma_{\mathrm{M}}(\vec{\hat{k}},\vec{\hat{p}})=  (\vec{\hat{k}}\times\vec{\hat{p}})\cdot\boldsymbol{\sigma}$ and $ \Gamma_{\mathrm{R}}(\vec{\hat{p}})=( \boldsymbol{\sigma} \times \vec{\hat{p}}) \cdot \hat{\vec{z}}$ are the interaction vertices for the Mott and Rashba scattering respectively. The dimensionless vertex strength $\lambda_{xy}$ and $\lambda_{z}$ can be treated as phenomenological parameters that parametrize the theory in different disorder regimes. For our microscopic model  (at zero temperature),  these parameters take the following values:  $\lambda_{xy}=p_{F}\lambda_{\perp}^2/8\xi$ and $\lambda_{z}=p_{F}^2\lambda_{\parallel}^{2} /4$, where $p_F$ is the Fermi momentum. In order to capture skew scattering and quantum interference effects one needs to go beyond the Gaussian  approximation~\cite{mirco2016crossover,mirco2016quantum} and consider up to the third moment in the distribution of the random potential:
\begin{align}
\langle V_{\vec{p}} \rangle &= 0 \\ 
\langle V_{\vec{p}}V_{\vec{q}}\rangle &=  n_{s} \, v_{0}^2 \, \delta(\vec{p}+ \vec{q}), \\ 
\langle V_{\vec{p}}V_{\vec{q}} V_{\vec{k}}\rangle &=  n_{s} \, v_{0}^3 \, \delta(\vec{p}+ \vec{q}+\vec{k}).
\end{align}
Here, $v_0$ is the strength of the impurity potential while $n_{s}$ has dimension of inverse area and it is usually identified with the impurity density. However, since in our model the electrons are scattered by a random -- albeit smooth --  potential, it is difficult to clearly identify $n_s$ as an impurity density. In what follows, $n_{s}$ should be understood as parametrizing the smoothness of the disorder landscape, depending itself on microscopic parameters such as the density of donors, the distance between the doping layer and the 2DEG, and the Thomas-Fermi screening length~\cite{glazov2010two}.
\section{Quantum Boltzmann Equation} \label{sec:QBE}
\begin{figure*}[t] 
\includegraphics[width=0.8\textwidth]{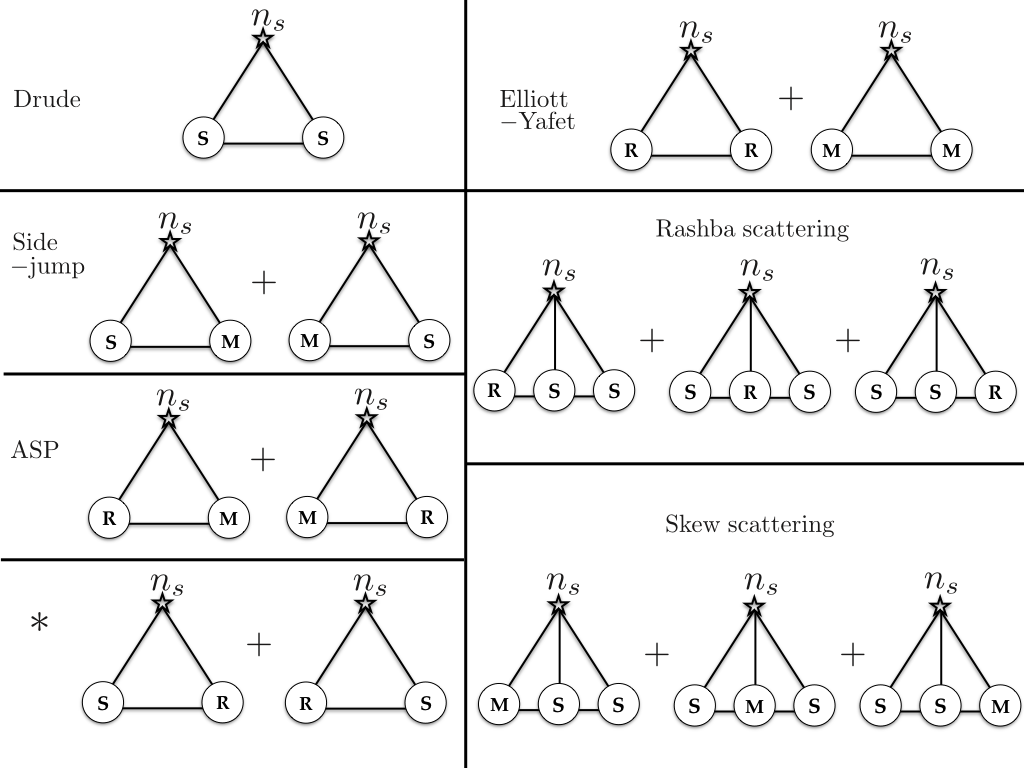}
\caption{Self-energy diagrams used in the evaluation of the collision integrals. The self-energy diagrams consist of three types of vertices: S, M, R. Note that the diagrams that contain different vertices lead to intriguing spin-charge conversion mechanism. In this sense, the conversion between spin and charge results from quantum interefence between different components of the spin-orbit coupling potentials. Neglecting diagrams containing the $\Gamma_{\mathrm{R}}$ vertex, we recover the results of Ref.~\onlinecite{shen2014theory,raimondi2012su2}. Note that the $*$ diagram vanishes for a non-polarized spin fermi liquid.}\label{fig:self_energy}
\end{figure*}

In this section we obtain the quantum Boltzmann equation describing coherent spin transport by means of the SU$(2)$ Schwinger-Keldysh formalism developed in Refs.~\onlinecite{raimondi2012su2,shen2014theory} [see also the appendix~\ref{a:SE} for details]. Under the influence of an electric ($\vec{E}$) and magnetic field ($\mathcal{H}^a$), the (matrix) distribution function $n_{\vec{p}}(\vec{r},t)$ satisfies the following equation:
\begin{align}
\left( \nabla_{t} n_{\vec{p}} +  \frac{\vec{p}}{m} \cdot   \nabla_{\vec{r}} n_{\vec{p}}  \right) +\frac{1}{2}\left\{\vec{F}_{\vec{p}} , \partial_{\vec{p}} n_{\vec{p}} \right\} 
= I[n_{\vec{p}}]. \label{eq:QBE_general}
\end{align}
Here $\vec{F}_{\vec{p}} =   \vec{E}+\tfrac{\vec{p}}{m} \times  \, \boldsymbol{\mathcal{B}} $ is the force acting on the electrons moving with velocity $\vec{p}/m$, and  $\boldsymbol{\mathcal{B}} = 8 m^2 \alpha^2 \boldsymbol{\hat{z}} \sigma^z$ is the effective ``magnetic field'' induced by the uniform Rashba SOC; $\nabla_{\vec{r}}$ and $\nabla_{t}$ are covariant derivatives accounting for spin precession induced by the SOC and the magnetic fields~\cite{raimondi2012su2,shen2014theory}:
\begin{align}
\nabla_{t}n_{\vec{p}}(\vec{r},t)=&\partial_t n (\vec{p},\vec{r},t)- i [(-\gamma \, \mathcal{H}^a \, \sigma^a), n_{\vec{p}}(\vec{r},t)], \\
\nabla_{r} n_{\vec{p}}(\vec{r},t)=& \partial_{\vec{r}} n(\vec{p},\vec{r},t) + i [ \, \boldsymbol{\mathcal{A}}^a \sigma^a, n_{\vec{p}}(\vec{r},t)],
\end{align}
where $\gamma$ is the gyromagnetic ratio. The left hand side of Eq.~\eqref{eq:QBE_general} describes the drift and diffusion of spin and charge due to both  the uniform SOC and the external  field. The right hand-side of Eq.~\eqref{eq:QBE_general} is the so-called collision integral.  We assume that the disorder induced, spin-orbit coupling strength is weak ($\lambda_{xy}\sim \lambda_{z} \ll 1$) and therefore omit terms that are proportional to third and higher order in $\lambda_{xy},\lambda_z$, and $v_{0}^{3} \, \lambda_{xy}, v_0^{3}\lambda_{z}$. In the standard, semiclassical Boltzmann equation, $I[n_{\vec{p}}]$ is assumed to be independent~\cite{KohnLuttingerBTE} of $\vec{E}$. However, this assumption neglects quantum interference effects between the electric field and the SOC potential. The side-jump contribution to spin-Hall effect precisely arises as a quantum mechanical correction to the  velocity operator~\cite{Nagaosa_RevModPhys.82.1539} $\vec{p}/m$. For these reasons, terms in the collision integral $I[n_{\vec{p}}]$ up to linear order in the components of electric field $E \, \lambda_z$ and $E \, \lambda_{xy}$ must be retained. Furthermore,  in the derivation of the right hand-side of Eq.~\eqref{eq:QBE_general},  we assumed a weak \emph{uniform} SOC strength, that is, $\alpha \ll  v_{F}$, where $v_F = p_F/m$ is the Fermi velocity. Therefore, we neglect any corrections to $I[n_{\vec{p}}]$ arising from the uniform SOC potential ($\alpha$ or $\boldsymbol{\mathcal{A}}$). Within the above approximations, we find that the collision integral can be associated with \emph{seven} distinct classes of self-energy diagrams shown in Fig.~\ref{fig:self_energy}. In the absence of an external magnetic field (i.e. for $\boldsymbol{\mathcal{H}}=0$), and for a uniform external electric field $\vec{E}$, the  quantum Boltzmann equation in the steady state  takes the form:
\begin{align} 
  \ i &\big[ e \, \boldsymbol{\mathcal{A}}^a \sigma^a \cdot \frac{\vec{p}}{m}\, ,\, n_{\vec{p}} \big]
 + \frac{1}{2}\left\{ e \vec{E}+  \vec{p} \times ( \hat{\vec{z}}  \omega_c  \sigma^z )   , \partial_{\vec{p}} n_{\vec{p}}\right\}  \nonumber \\
=  &\, I_{\mathrm{D}}[n_{\vec{p}}]+ I_{\mathrm{EY}}[n_{\vec{p}}]+I_{\mathrm{RS}}[n_{\vec{p}}] + I_{\mathrm{SS}}[n_{\vec{p}}]+ I_{\mathrm{SJ}}[n_{\vec{p}}] \nonumber \\&+ I_{\mathrm{ASP}}[n_{\vec{p}}].
\label{eq:QBE}
\end{align}
%
%
Note that the covariant space derivative does not vanish even for the uniform steady state where $n_{\vec{p}}$ is independent of $\vec{r}$ but contributes a spin-precession term induced by the uniform SOC~\cite{shen2014theory}. Here we have parametrized the strength of the effective SOC magnetic field using $\omega_c =  |\boldsymbol{\mathcal{B}}|/m=8 \, m \, \alpha^2 $, which is the cyclotron frequency induced by the uniform SOC. The collision integrals on the right hand side of Eq.~\eqref{eq:QBE} correspond respectively to: Drude relaxation, Elliott-Yafet relaxation, Rashba Scattering, skew scattering, side-jump and anisotropic spin precession scattering. Their evaluation  is described in Appendix~\ref{a:collision}.
\section{Linear Response Matrix}  \label{sec:result1}
As discussed in e.g. Refs.~\onlinecite{chunli2016graphene,chunli2016direct}, in the steady state the quantum Boltzmann equation can be solved using the following  \textit{ansatz}:
\begin{equation} \label{eq:ansatz}
n_{\vec{p}} = n_{\mathrm{FD}}\left[\epsilon_p -   \boldsymbol{p} \cdot \boldsymbol{v}_{c} - ( \, ( \boldsymbol{p} \cdot \boldsymbol{v}_{s}) \, \boldsymbol{\hat{n}}_{1} + h_0 \, \boldsymbol{\hat{n}}_{0})  \cdot \boldsymbol{\sigma}\right].
\end{equation}
Here $ n_{\mathrm{FD}}(\epsilon)  = \left[e^{(\epsilon-\mu)/k T}+1\right]^{-1}$ is the Fermi-Dirac distribution function, $\epsilon_p = p^2/2m$ the electron kinetic energy, $\boldsymbol{v}_{c}$ ($\boldsymbol{v}_{s}$)  the drift velocity of the charge (spin) degrees of freedom, $h_0$ is proportional to the magnitude of the magnetization, and $\boldsymbol{\hat{n}_0}$ and $\boldsymbol{\hat{n}_1}$ are respectively the directions of the magnetization and the spin current.
 Our ansatz amounts to solving the Boltzmann equation with an expansion in circular
harmonics of the Fermi surface deformation \cite{landau_lifshitz_sm1}. We are interested in evaluating the non-equilibrium spin polarization $\boldsymbol{M} = (M^x, M^y, M^z)$, the charge current density  $\boldsymbol{J} = (J_x, J_y)$  and  the spin current density $\boldsymbol{\mathcal{J}}^{a} = (\mathcal{J}_x^a, \mathcal{J}_y^a)$  ($a=x,y,z$ is the spin orientation). At zero temperature, these observables are related to the parameters of the \textit{ansatz}~\eqref{eq:ansatz}  as follows:
\begin{align}
M^a =&\frac{1}{2} \sum_{\vec{p}} \mathrm{Tr} \left[  \sigma^a n_{\vec{p}} \right] = N_0 \, h_{0} \: \hat{n}_{0}^a,\\
J_i = &\frac{1}{2} \sum_{\vec{p}}  \mathrm{Tr}  \left[  \frac{p_{i}}{m} n_{\vec{p}} \right]  =N_0 \, \epsilon_F \, \frac{(v_{c})_i}{2},\\
\mathcal{J}^{a}_i =&\frac{1}{2} \sum_{p}  \mathrm{Tr} \left[   \sigma^a \frac{p_{i}}{m} n_{\vec{p}} \right] =N_0 \, \epsilon_F \, \frac{(v_{s})_i \hat{n}_1^a}{2},
\end{align}
were $N_0=\sum_{p} \delta(\epsilon_p -\mu)=m/2\pi$ is the density of states of the 2DEG. In order to make contact with the results of Ref.~\onlinecite{chunli2016direct},  we shall measure the spin and charge currents in the same units and we rescale the magnetization by defining $\boldsymbol{\mathcal{M}} = v_{F} \, \boldsymbol{M}$.  Substituting Eq.~\eqref{eq:ansatz} into Eq.~\eqref{eq:QBE} and setting $\vec{E}=E_x \, \hat{\vec{x}}$, we finally obtain the following linear response relations:
\begin{multline} \label{eq:transport}
\begin{pmatrix}
J_x \\ \mathcal{J}_y^z \\ \mathcal{M}^y \end{pmatrix}
=
\begin{pmatrix}
0 & \theta_{\mathrm{sH}} &  \tau_\mathrm{D} \alpha_{\mathrm{asp}} \\
-\theta_{\mathrm{sH}} & 0 &  \tau_\mathrm{D} \alpha_{\mathrm{R}} \\
\tau_{\mathrm{EY}}^y \alpha_{\mathrm{asp}} 
& - \tau_{\mathrm{EY}}^y \alpha_{\mathrm{R}} & 0 
\end{pmatrix}
\begin{pmatrix}
J_x \\
\mathcal{J}_y^z \\
\mathcal{M}^y
\end{pmatrix}
\\+ 
\begin{pmatrix}\sigma_{\mathrm{D}} \, E_x \\ \sigma^{SJ}_{yx} \, E_x \\ 0 \end{pmatrix}.
\end{multline}
It is easy to see that under the influence of an electric field $\vec{E}$, the 2DEG responds in three different ways: a longitudinal charge current $\vec{J} = J_x \: \vec{\hat{x}}$  and  a transverse spin current $\boldsymbol{\mathcal{J}}^z = \mathcal{J}_{y}^{z} \: \vec{\hat{y}}$  and a non-equilibrium spin-polarization $\boldsymbol{\mathcal{M}} = \mathcal{M}^y \: \boldsymbol{\hat{y}}$. The direction of flow and spin-polarization of these responses is determined by the direction of the external electric field $\vec{E}$ and the symmetry of the system.

The first term on the right hand side of Eq.~\eqref{eq:transport} describes the coupling between responses (induced by both intrinsic and extrinsic SOC) while the second term describes their relation to the applied driving field; this is in turn characterized by the Drude conductivity $\sigma_{\mathrm{D}}$ and the contribution of the spin-Hall conductivity arising from the side-jump mechanism,  $\sigma^{SJ}_{yx}$:
\begin{equation}
\sigma_{\mathrm{D}}= \frac{n  \, \tau}{m}  \; ,\; 
\sigma^{SJ}_{yx} = n \, \lambda_{z},
\end{equation}
where $n=\sum_p f_{\mathrm{FD}}(\epsilon_p)$ is the electron density. Most importantly, the coupling matrix in Eq.~\eqref{eq:transport} is characterized by three ``conversion rates'' between different responses (i.e.~$\{  J_x , \mathcal{J}_y^z,  \mathcal{M}^y  \}$) and by two relaxation rates.The two relaxation rates are the elastic (Drude) scattering rate and the Elliott-Yafet spin scattering rate. The elastic (Drude) scattering rate and the (anisotropic) Elliott-Yafet scattering rate are given by:
\begin{align}
&\frac{1}{\tau_{\mathrm{D}}}= 2\pi \, n_s \, v_{0}^2 \, N_0 \;  , \;
\frac{1}{\tau_{\mathrm{EY}}^{z}} =\frac{4 \, \lambda_{xy}^2 }{\tau_{\mathrm{D}}} \\
&\frac{1}{\tau_{\mathrm{EY}}^{x}}=\frac{1}{\tau_{\mathrm{EY}}^{y}}=\frac{\left(6 \, \lambda_{xy}^2 + \lambda_{z}^2 \right)}{\tau_{\mathrm{D}}}.
\end{align}
The other three conversion rates (or  ratios) are
\begin{align}
\alpha_{\mathrm{R}}&=p_{F} \left( \alpha +  \delta \alpha_{\mathrm{R}} \right) , \label{eq:alpha_r}
\\
\alpha_{\mathrm{asp}}&=-4 \, \lambda_{xy} \, \lambda_{z} \, \tau_{\mathrm{D}}^{-1}, \label{eq:ASP} \\
\theta_{\mathrm{sH}}&= 2\pi \, v_0 \, N_0 \, \lambda_z + \omega_c \, \tau_{\mathrm{D}}. \label{eq:SHE}
\end{align}
$\alpha_{\mathrm{R}}$ is the conversion rate between the (macroscopic) spin-current $\mathcal{J}_{y}^{z}$ and the non-equilibrium magnetization $\mathcal{M}^y$. In addition to the usual contribution from the uniform Rashba SOC (i.e.~$\alpha \, p_F$), it also receives a renormalization coming from the  disorder  induced energy shift $\delta\alpha_{\mathrm{R}}$.
 Diagrammatically, $\delta\alpha_{\mathrm{R}}$ arises from the (third order Born approximation) Rashba scattering diagram and part of the ASP diagram in Fig.~\ref{fig:self_energy}, see Appendix B for more information. This impurity induced precession was also found in Ref.~\onlinecite{chunli2016direct} as a result of self-energy correction (i.e.~the part of the collision integral that is linear order in the $T$-matrix).
Similarly, $\alpha_{\mathrm{asp}}$ in Eq.~\eqref{eq:ASP} is the anisotropic spin precession scattering rate, inducing conversion between the (macroscopic) charge current $J_x$ and the magnetization $\mathcal{M}^y$, see Fig.~\ref{fig:self_energy}. 
The spin Hall angle -- Eq.~\eqref{eq:SHE} -- contains both the skew scattering and intrinsic contribution~\cite{shen2014theory}. The latter arises from the uniform Rashba SOC and is proportional to the  ``cyclotron frequency''  $\omega_c  = 8 \, m \, \alpha^2$. The intrinsic and skew scattering contributions arise from the non-equilibrium part of the distribution function~$n_{\vec{p}}$.  However, the side-jump contribution to the spin-Hall conductivity involves the equilibrium distribution. The difference is  reflected in Eq.~\eqref{eq:transport}: The side-jump couples the  spin current $\mathcal{J}_y^z$ directly to the \textit{electric field} via $\sigma^{SJ}_{yx}$. On the other hand,  the intrinsic and skew scattering mechanisms couple $\mathcal{J}_y^z$  to the \textit{charge current}. Nevertheless, this distinction is not important when solving \eqref{eq:transport} for the total spin Hall conductivity.  Note that the uniform SOC (i.e.~non-abelian gauge field $\alpha$) will only quantitatively change the spin Hall angle $\theta_{\mathrm{sH}}$ and the Rashba conversion rate $\alpha_{\mathrm{R}}$, but it will leave the form of the linear response equation (Eq.~\ref{eq:transport}) unchanged.

The linear response matrix, Eq.~\eqref{eq:transport},  has been obtained within the SU(2) Schwinger-Keldysh formalism assuming a weak  (but smooth) disorder potential.  An almost identical result has been also obtained within the Kohn-Luttinger formalism developed in Ref.~\onlinecite{chunli2016direct}, under the assumption that the impurity density is small (but for arbitrarily strong single-impurity potential). Notice that in the Kohn-Luttinger approach,  the side-jump contribution is absent to leading order in the impurity density~\cite{chunli2016direct}. However, the strong similarities between the  transport theories resulting from two very different microscopic models suggest that the quantum interference effects induced by SOC disorder potentials and the electric field are fairly universal transport phenomena.
\section{Current-induced spin polarization and spin current} \label{sec:result2}
In this section, we use  the linear response matrix equation, Eq.~\eqref{eq:transport}, to discuss the phenomena of current-induced spin polarization and current-induced spin current. Invert the matrix in Eq.~\eqref{eq:transport} and solve for $\mathcal{J}_{y}^z$ and $\mathcal{M}^y$ as a function of $E_x$ and considering that the spin-charge conversion rates (i.e.~$\alpha_{\mathrm{asp}} \, \tau_{\mathrm{D}},\theta_{\mathrm{sH}},\alpha_{\mathrm{R}}\, \tau_{\mathrm{D}}$) are typically small, we obtain the following results within linear response theory:
\begin{equation} \label{eq:sigmas}
\mathcal{J}_{y}^{z} = \left(\sigma_{\mathrm{sH}} +\sigma_{\mathrm{ind}} \right) E_x  \, , \,
\mathcal{M}^{y} = \left(\sigma_{\mathrm{DMC}} +\sigma_{\mathrm{EE}} \right) E_x. 
\end{equation}
These two equations account for the charge-induced spin current and spin polarization, respectively. The  ratio of the spin current (spin polarization) to the electric field corresponds to the conversion efficiency of the  material.

The current-induced-spin current receives direct and  indirect contributions: The direct contribution is proportional to the spin-Hall conductivity $\sigma_{\mathrm{sH}}$ and  arises from the SHE, which converts  the charge current $J_x$ into the spin-current $\mathcal{J}^z_y$. An indirect contribution to the spin current is proportional to $\sigma_{\mathrm{ind}} $ and arises in a two-step process in which  $J_x \rightarrow \mathcal{M}^y$ -- by virtue of ASP scattering -- followed by a process in which $M^y \rightarrow \mathcal{J}_{y}^z$, by virtue of  the precession induced by the Rashba field~\cite{Kashen2014micro}. The spin conductivities are given by
\begin{align}
&\sigma_{\mathrm{sH}}=- \omega_c \, \tau_{\mathrm{D}} \, \sigma_{\mathrm{D}}-  v_0  \, \lambda_z \, p_{F}^2 \, n \, \tau_{\mathrm{D}} - n \, \lambda_z   \\
&\sigma_{\mathrm{ind}}= (\alpha_{\mathrm{R}} \, \tau_{\mathrm{D}} )(\alpha_{\mathrm{asp}} \, \tau_{\mathrm{EY}}^{y}) \,\sigma_{\mathrm{D}} 
\end{align}

Similarly, the current-induced spin polarization also receives  direct and indirect contributions. The direct contribution is characterized by the direct-magneto electric coupling $\sigma_{\mathrm{DMC}}$ and it arises from  a  conversion $J \rightarrow \mathcal{M}$ process. The indirect contribution arises from the Edelstein effect~\cite{Edelstein1990233,amin2016the} $\sigma_{\mathrm{EE}}$ and it is characterized by the conversion sequence~$J_x \rightarrow \mathcal{J}_{y}^z \rightarrow \mathcal{M}^y$. Their explicit form is given by 
\begin{align} \label{eq:DMC}
\sigma_{\mathrm{DMC}}= (\alpha_{\mathrm{asp}} \tau_{\mathrm{s}}^{y}) \sigma_{\mathrm{D}}   \; , \;
\sigma_{\mathrm{EE}}= (\alpha_{\mathrm{R}} \tau_{\mathrm{s}}^{y})\sigma_{\mathrm{sH}}. 
\end{align}
The total spin relaxation time $\tau_{s}^{y}$ consists of the Elliott-Yafet and D\'yakonov-Perel mechanisms:
\begin{equation}
\frac{1}{\tau_{s}^{y}}=\frac{1}{\tau_{\mathrm{EY}}^{y}}+ \frac{1}{ \tau_{\mathrm{DP}}}.
\end{equation}
Here $\tau_{\mathrm{DP}}^{-1}=\tau_{\mathrm{D}}(\alpha_{\mathrm{R}}^2-\alpha_{\mathrm{asp}}^2)$. Note that $\sigma_{\mathrm{DMC}}$ arises from the ASP scattering. We would like to stress that this is different from the microscopic origin of the Eldestein effect~\cite{Kashen2014micro}, for which the non-equilibrium spin polarization arises via the conversion sequence: $J_x \rightarrow \mathcal{J}_{y}^z \rightarrow \mathcal{M}^y$.
 Note that the figures of merit of current-induced spin polarization,  $\sigma_{\mathrm{EE}}$ and $\sigma_{\mathrm{DMC}}$, are proportional to the total spin relaxation time. In non-uniform systems, ASP modifies the spin continuity equation as follow:
\begin{equation}
[\nabla_{t} \,m]^{a}+ [\nabla_{i}{\mathcal{J}}_i] ^a 
= -\frac{m^a}{\tau_{\mathrm{EY}}^a} + \alpha_{\mathrm{asp}} \,\epsilon^{ajz} \, J_j. 
\end{equation}
which can be derived from the Quantum Boltzmann equation as explained in Ref.~\onlinecite{chunli2017anomalous}. Here $\epsilon^{ijk}$ is the Levi-Civita antisymmetric tensor. Note that ASP scattering is a form of direct magnetoelectric coupling (DMC) since it couples spin density $m^a$ to the electric current $J_i$ (hence electric field) directly in the spin continuity equation without resorting to any constitutive relations. 

\section{DIscussions of experiments} \label{sec:exp}
The Spin Hall effect (SHE) and the current-induced spin polarization (CISP) are ubiquitous transport phenomena that have been observed in Ref. \onlinecite{kato2004observation,valenzuela2006direct,ahmet2015enhanced} and Ref.~\onlinecite{sanchez2013spin,kato2004current,song2016spin,song2017observation}, respectively.
Although their relative contributions to the overall spin-charge conversion will depend on the microscopic details of the materials, both effects can occur together and couple with each other on symmetry grounds \cite{chunli2016direct}. In order to differentiate CISP from SHE, Ref.~\onlinecite{sih2005spatial} used optical Kerr-rotations to study the direction of the spin polarization and its spatial accumulations. In addition to optical methods, some authors of the present article also proposed an all-electrical experiment~\cite{chunli2017anomalous} in order to differentiate SHE from CISP, based on the theory first developed in Ref.~\onlinecite{abanin2009nonlocal}. 

Recently,  J.~Bindel \textit{et.~al} reported on the fluctuations of Rashba SOC in InSb inversion layer~\cite{bindel2016probing}. 
From the Supplementary Material of Ref.~\onlinecite{bindel2016probing}, the elastic scattering time $\tau_{\mathrm{D}}\sim 200\mathrm{ps}$ and the
\emph{giant} uniform Rashba SOC strength $\alpha=1.2\mathrm{eV}\AA$. 
From Fig.~3f of Ref.~\onlinecite{bindel2016probing}, we can estimate the correlation between fluctuating Rashba strength and the 2D potential energy,  $\lambda_{\perp}^2/ \xi \sim 12\AA$. Therefore, $\alpha_{\mathrm{asp}}^{-1}\sim 10^{4} \tau_{\mathrm{D}}$, $\tau_{\mathrm{EY}}\sim 10^{3} \tau_{\mathrm{D}}$, meaning that we are in the limit $\alpha_{\mathrm{asp}}^{-1} > \tau_{\mathrm{EY}} \gg \tau_{\mathrm{D}}$. 
In terms of spin-charge conversion efficiencies, we found $\sigma_{\mathrm{DMC}}/\sigma_{\mathrm{EE}} =\alpha_{\mathrm{asp}}/(\alpha_{\mathrm{R}}\theta_{\mathrm{sH}}) \sim 10^{-3}$ due to the large spin Hall angle arising from the giant uniform Rashba SOC,
 i.e. the spin Hall angle is dominated by intrinsic contribution $\theta_{\mathrm{sH}}\sim \omega_c \tau_{\mathrm{D}} \sim 1$. Hence, the SHE dominates the CISP in Ref.~\onlinecite{bindel2016probing}.

The  ratio $\sigma_{\mathrm{DMC}}/\sigma_{\mathrm{EE}}$ can be enhanced in 2D systems with small or vanishing uniform Rashba SOC like symmetrically doped 2DEG \cite{glazov2010two} or adatoms decorated graphene \cite{chunli2016direct}. This is because in our theory, the DMC arises from the extrinsic mechanism (i.e.~ASP scattering), whereas both SHE and the conversion between spin-current and spin density receive contributions from both extrinsic and intrinsic mechanisms. It is interesting to understand how can DMC occur in 2D systems without relying on disorder potentials that break spatial inversion symmetry.

Note that our calculations are  presented for the zero temperature case. At finite temperature, and if the system can support resonant scattering (as in the case of adatoms decorated graphene) finite temperature effects will broaden the line width and decrease the amplitude of the spin Hall conductivity, as described in Ref.~\onlinecite{hy2015extrinsic}. We expect the same finite temperature behaviour to occur in  $\sigma_{\mathrm{DMC}}$ as well since the broadening of line width and decrease in peak amplitude results from averaging on the number of states near the Fermi surface, and is largely independent of the scattering mechanism.
However, in a 2DEG resonant scattering is difficult to observe due to the lack of energy dependence of its density of states. From Eq.~\ref{eq:DMC},  $\sigma_{\mathrm{DMC}}=\alpha_{\mathrm{asp}}\tau_{\mathrm{s}}^{y}\sigma_{D}$ and Eq.~\ref{eq:ASP} $\alpha_{\mathrm{asp}}=-4\lambda_{xy}\lambda_{z}\tau_{\mathrm{D}}^{-1}$,  we found that the temperature dependence of DMC follows from the temperature dependence of the total spin relaxation time: $\sigma_{\mathrm{DMC}}(T) \propto\tau_{\mathrm{s}}^{y}(T)$. 
The temperature dependence of spin relaxation time depends on the microscopic details of a particular system (e.g.~mobility, symmetrically or asymmetrically doped quantum well). For example, Ref.~\onlinecite{han2011temperature} reported D'yakonov-Perel mechanism as the dominant spin relaxation channel in their experiment and it has interesting non-monotonic temperature dependence. Therefore, DMC would still be observable at temperature where spin-relaxation time does not vanish.

\section{Summary and Outlook}

In this work we have obtained the linear response of a two-dimensional electron gas under the influence of both intrinsic SOC and a smooth disorder SOC landscape. In particular, by  accounting for both spin-conserving (Mott) and spin non-conserving (Rashba) scattering processes,  we have found that the quantum interference between them gives rise to the anisotropic spin precession scattering first found in an earlier  study of spin-transport in graphene decorated with adatoms~\cite{chunli2016direct}. The anisotropic spin precession scattering is a form of direct magneto electric coupling which gives contributions to both the current-induced spin polarization and the current-induced spin current.  Our results suggest that  this mechanism, which describes the polarizing effect of the disorder SOC potential should be a rather universal  phenomenon in disordered two-dimensional metals lacking inversion symmetry. 

\section*{Acknowledgements}
We gratefully acknowledge Roberto Raimondi for kindly delivering a series of lectures on the SU$(2)$-covariant Schwinger-Keldysh formalism after the workshop ``Recent Progress in Spintronics of 2D Materials'' held  at the National Center for Theoretical Sciences in Taiwan. C.H and M.A.C acknowledge support from the Ministry of Science and Technology (Taiwan) under contract No.~NSC~102-2112-M-007-024-MY5 and Taiwan's National Center of Theoretical Sciences (NCTS).  C.H. also acknowledges support from the Singapore National Research Foundation grant No.~NRFF2012-02, and from the Singapore Ministry of Education Academic Research Fund Tier 2 Grant No. MOE2015-T2-2-008.  M. M. thanks C. Verma for his hospitality at the Bioinfomatics Institute in Singapore where this work was initiated. C. H. gratefully acknowledges the hospitality of the Donostia International Physics Center.

\appendix

\section{Self-energy diagrams and collision integrals}\label{a:SE}
In the Schwinger-Keldysh transport formalism~\cite{rammer1986rmp,rammer_book}, the collision integral reads
\begin{align}
I=  \int  \frac{d\epsilon}{4\pi } \left( -\Sigma^R G^{K}
-\Sigma^{K} G^{A} +G^{R} \Sigma^{K}  +G^{K}\Sigma^{A}\right), \label{eq:kelcollint}
\end{align}
where $G^{R}$,$G^A$ and $G^K$ are the retarded, advanced and Keldysh components of the  Green's function, respectively. Similarly, $\Sigma^{R}$ , $\Sigma^A$ and $\Sigma^K$ are the retarded, advanced and Keldysh components of the self-energy. The disorder self-energy is a four by four matrix  in Keldysh and spin space. In order to evaluate the collision integral, we use the quasi-particle approximation, which approximates
\begin{equation}
G^R(\epsilon,\vec{p})=\frac{1}{\epsilon-\epsilon_{p} +i \, \delta} \; ;\; 
G^A(\epsilon,\vec{p})=\frac{1}{\epsilon-\epsilon_{p} -i \, \delta} \label{eq:approx1}
\end{equation}
\begin{equation}
G^K (\epsilon,\vec{p})= -2\pi \, i \, \delta(\epsilon- \epsilon_p) \, (1-2n_{\vec{p}}). \label{eq:approx2}
\end{equation}
In other words,  Eq.~\eqref{eq:approx1} ignores the disorder-induced broadening of the spectral function, whereas  Eq.~\eqref{eq:approx2} assumes the existence of a local equilibrium distribution function. The leading order corrections to the collision integral arise from the electric field and are proportional to $E\lambda_{xy}$ and  $E\lambda_{z}$.  We will neglect any higher order corrections in the electric field. Since the uniform Rasba SOC is  weak and the self-energy is at least second order in the impurity scattering potentials, we will also neglect the correction due to the uniform Rashba SOC potential in  the evaluation of the collision integrals, i.e. all collision integrals are zeroth order in the uniform Rashba potential strength $\alpha$. 
\section{Collision integrals} \label{a:collision}
In this appendix, we  provide the most important details of the computation of the different contributions  to the collision integral, Eq.~\eqref{eq:QBE}. From here on, we shall use the short-hand notation $p=(\epsilon, \vec{p})$, $p'=(\epsilon, \vec{p}')$, etc. Note that the energy $\epsilon$ is unchanged since  scattering with the disorder potential is elastic.
The relevant Feynman diagrams are shown in Fig.~\ref{fig:self_energy}.
\subsection{Drude relaxation: Scalar-scalar self-energy}
For this diagram, after disorder average, the Keldysh self-energy matrix is
\begin{equation}
\tilde{\Sigma}(p)=  n_{i}v_{0}^2 \sum_{\vec{p'}} \tilde{G}(p'),
\end{equation}
where the tilde means that the (Green's) function is SU($2$) locally covariant. After inserting this result into Eq.~\eqref{eq:kelcollint}, the resulting collision integral yields the standard Drude relaxation term:
\begin{equation}
I_{\mathrm{D}}[n_{\vec{p}}] = 2\pi \, n_{i} \, v_0^2 \, \sum_{\vec{p}'} \, (n_{\vec{p}'} - n_{\vec{p}}) \, \delta(\epsilon_p - \epsilon_{p'}). \label{eq:coll_drude}
\end{equation}
The Drude term drives the relaxation of  the charge and spin current.
\subsection{Anisotropic spin precession scattering: Mott-Rashba self-energy}
The self-energy for anisotropic spin precession scattering is given by:
\begin{align}
\tilde{\Sigma}(p)= i \, n_{i} \, v_{0}^2 \, \lambda_{xy} \, \lambda_{z}  &\sum_{\vec{p'}} \bigg(
\Gamma_{\mathrm{M}}(\vec{p},\vec{p}') \, \tilde{G}(p')\, \Gamma_{\mathrm{R}}(\vec{p}+\vec{p}') \nonumber \\
&-\Gamma_{\mathrm{R}}(\vec{p}+\vec{p}') \, \tilde{G}(p') \, \Gamma_{\mathrm{M}}(\vec{p},\vec{p}')
 \bigg).
\end{align}
Notice that it corresponds to a quantum interference process between the Mott and Rashba-type components of the
SOC disorder potential.  The resulting collision integral can be split as the sum $I_{\mathrm{ASP}}=I_{\mathrm{ASP}}^{0}[n _{\vec{p}}]+I_{\mathrm{ASP}}^{1}[n _{\vec{p}}]$. 
\begin{align}
&I_{\mathrm{ASP}}^{0}[n _{\vec{p}}]=2\pi \, n_{i} \,  v_0^2 \, \lambda_{xy} \, \lambda_{z} \,  i \, \sum_{\vec{p}'}  \, \delta(\epsilon_p - \epsilon_{p'})   \nonumber \\
&\times \, \bigg( \Gamma_{\mathrm{R}}(\vec{p}+\vec{p}') \, n_{\vec{p'}} \, \Gamma_{\mathrm{M}}(\vec{p},\vec{p}') -\Gamma_{\mathrm{M}}(\vec{p},\vec{p}') \, n_{\vec{p}'} \, \Gamma_{\mathrm{R}}(\vec{p}+\vec{p}') \nonumber \\
&+\frac{1}{2}\big\{  \left[ \Gamma_{\mathrm{M}}(\vec{p},\vec{p}'), 
 \Gamma_{\mathrm{R}}(\vec{p}+\vec{p}')\right]  , n_{\vec{p}}\big\}
 \bigg)
\end{align}
Using the  \textit{ansatz} \eqref{eq:ansatz} introduced in Sec.~\ref{sec:result1}, we obtain
\begin{align}
\sum_{\vec{p}}I_{\mathrm{ASP}}^{0}[n_{\vec{p}}]=&\sum_{\vec{p}}I_{\mathrm{ASP}}[\delta(\epsilon_p -\mu)\vec{v}_c \cdot \vec{p}] \nonumber \\
=&\frac{4 \, \lambda_{xy} \, \lambda_{z}}{\tau_{\mathrm{D}}v_F}\left( \sigma^y J_x -\sigma^x J_y \right)
\end{align}
The second part of the collision integral resembles a precession term that can be absorbed into the right-hand side of the Boltzmann equation (i.e. the non-dissipative part):
\begin{align}
I_{\mathrm{ASP}}^{1}[n _{\vec{p}}]=2\pi^2 i  \lambda_{z}\lambda_{xy}n_{i}v_{0}^2  N_{0}^{2} p_{F}^{-3} \kappa(D,\epsilon_{F})  \left[  \Gamma_{R}(\vec{p}),n_{\vec{p}}\right].
\end{align}
Here $\kappa(D,\epsilon_{F})$ is a parameter depending on the energy cut-off (bandwidth) $D$ and Fermi energy $\epsilon_{F}$:
\begin{equation}
\kappa(D,\epsilon_{F})= \mathcal{P} \int_{0}^{D} \frac{d \epsilon}{\pi} \, \left[ \frac{\epsilon}{\epsilon_F -\epsilon} \right]
\end{equation}
\subsection{The $\star$ diagram: Scalar-Rashba self-energy}
In this case the self-energy  is
\begin{align}
\tilde{\Sigma}(p)=& - n_{i} \, v_{0}^2 \, \lambda_{xy} \, \sum_{\vec{p'}} \bigg(  \{ \tilde{G}(p')\, , \,\Gamma_{\mathrm{R}}(\vec{p}+\vec{p}') \} \nonumber \\
&+   i e \left[ \partial_{\epsilon} \tilde{G}(p'), (\vec{E}\times \sigma )\cdot z\right]  \bigg)
\end{align}
The related collision integral is $I[n_{\vec{p}}]=I_{\star}^{0}[n_{\vec{p}}]  + I_{\star}^{1}[n_{\vec{p}}]$ where 
\begin{align}
I_{\star}^{0}[n_{\vec{p}}]=&2\pi \, n_{i} \, v_0^2 \, \lambda_{xy} \, \sum_{\vec{p}'} \, \delta(\epsilon_p - \epsilon_p')  
\left\{  \Gamma_{\mathrm{R}}(\vec{p}'+\vec{p}),n_{\vec{p}'}  -  n_{\vec{p}}   \right\}  
\end{align}
\begin{align}
I_{\star}^{1}[n_{\vec{p}}]&= 2\pi \, n_{i} \, v_0^2 \, \lambda_{xy} \, \sum_{\vec{p}'}   
\partial_{\epsilon_p} \, \delta(\epsilon_p - \epsilon_p') \nonumber \\
&\times \, i \, \left[ (-e\vec{E}) \cdot (\boldsymbol{\sigma} \times \boldsymbol{\hat{z}}) ,  n_{\vec{p}'}  \right].
\end{align}
$I_{\star}^{1}[n]$ is proportional to $En_{i}$ and it contributes to the anomalous velocity. In the linear response regime, $I_{\star}^{1}[n_{\vec{p}}^{0}]=0$ for a spin-unpolarized ground state. Using the drift velocity \textit{ansatz},  $\sum_{p}I_{\star}^{0}[\delta n_{\vec{p}}]=\sum_{p} \vec	{p} I_{\star}^{0}[\delta n_{\vec{p}}]=0$, since $\Gamma_{R} \propto \cos \theta$ while $\delta n_{\vec{p}'}  - \delta n_{\vec{p}} \propto \sin \theta$, where $\theta=(\theta_p -\theta_{p'})/2$.
\subsection{Rashba-scattering: Rashba-scalar-scalar  self-energy}
For this diagram, after disorder average, the self-energy  reads:
\begin{align}
\tilde{\Sigma}(p) &= -n_{i} \, v_{0}^3 \, \lambda_{xy} \, \sum_{\vec{p}_a \vec{p}_b}
\bigg[ \Gamma_{\mathrm{R}}(\vec{p}+\vec{p}_a) \tilde{G}(p_a) \tilde{G}(p_b)  \nonumber \\
&+  \tilde{G}(p_a) \Gamma_{\mathrm{R}}(\vec{p}_a+\vec{p}_b)\tilde{G}(p_b) +  \tilde{G}(p_a)\tilde{G}(p_b) \Gamma_{\mathrm{R}}(\vec{p} +\vec{p}_b) \bigg]
\end{align}
The resulting collision integral is given by the following expression:
\begin{align}
I_{\mathrm{RS}}[n_{\vec{p}}] = 2\pi^2 \, n_{i} \, v_0^3 \, \lambda_{xy} \, p_{F}^{-1} \, N_{0}^2 \, i \, \left[ n_{\vec{p}} , \Gamma_{\mathrm{R}}(\vec{p}) \right] \label{eq:coll_Rashba_scatt}
\end{align}
where $N_0=\sum_p \delta(\epsilon-\epsilon_p)=m/2\pi$ is the density of states. Together with $I_{\mathrm{ASP}}^{1}$, they can be absorbed into the SOC gauge field $\mathcal{A}_{y}^{x}=2m\alpha$ on the right-hand side of the quantum Boltzmann equation and this leads to a renormalization of the parameter $\alpha \rightarrow  \alpha + \delta \alpha_{R}$ where
\begin{equation}
\delta \alpha_{R} = n_{i} \pi v_0^2 N_{0}^2 p_{F}^{-1}\lambda_{xy}  \left( v_0^2 - \lambda_{z} \kappa(D,\epsilon_{F}) \right).
\end{equation}
\subsection{Elliott-Yafet relaxation: Rashba-Rashba and Mott-Mott self-energy}
The self-energy leading to spin-relaxation by the Elliiot-Yafet mechanism 
is given by the following expression:
\begin{align}
\tilde{\Sigma}(p)=& n_{i} \, v_{0}^2 \, \lambda_{xy}^2 \, \sum_{\vec{p'}} \, \Gamma_{\mathrm{R}}(\vec{p}+\vec{p}') \, \tilde{G}(p') \Gamma_{\mathrm{R}}(\vec{p} + \vec{p}').
\end{align}
\begin{align}
\tilde{\Sigma}(p)=& -n_{i} \, v_{0}^2 \, \lambda_{z}^2 \, \sum_{\vec{p'}} \, \Gamma_{\mathrm{M}}(\vec{p},\vec{p}') \, \tilde{G}(p') \Gamma_{\mathrm{M}}(\vec{p}',\vec{p}).
\end{align}
The resulting collision integral is the Elliott-Yafet spin relaxation term coming from the Mott-vertex and the Rashba vertex $I_{\mathrm{EY}}[n_{\vec{p}}]=I_{\mathrm{EY}}^{M}[n_{\vec{p}}]+I_{\mathrm{EY}}^{R}[n_{\vec{p}}]$:
\begin{align}
I_{\mathrm{EY}}^{R}[n_{\vec{p}}] &= 2\pi  \, n_{i} \, v_0^2 \, \lambda_{xy}^2 \, \sum_{\vec{p}'}  \, \delta(\epsilon_p - \epsilon_p')  \Big[ \Gamma_{\mathrm{R}}(\vec{p}+\vec{p}') \, n_{\vec{p}'} \nonumber \\ 
&\times \,  \Gamma_{\mathrm{R}}(\vec{p}+\vec{p}')  -\frac{1}{2} \{ \Gamma_{\mathrm{R}}^2(\vec{p}+\vec{p}') \, ,\, n_{\vec{p}}\} \Big].
\label{eq:coll_EY1}\\
I_{\mathrm{EY}}^{M}[n_{\vec{p}}] &= 2\pi \, n_{i} \, v_0^2 \, \lambda_{z}^2 \, \sum_{\vec{p}'}  \delta	(\epsilon_p - \epsilon_p')  \Big[ \Gamma_{\mathrm{M}}(\vec{p}, \vec{p}') n_{\vec{p}'} \Gamma_{\mathrm{M}}(\vec{p},\vec{p}') \nonumber \\ & -\frac{1}{2} \{ \Gamma_{\mathrm{M}}^2(\vec{p},\vec{p}') \, ,\, n_{\vec{p}}\} \Big].
\label{eq:coll_EY2}
\end{align}

\subsection{Side-jump and swap current: Mott-scalar self-energy}
The side-jump self-energy diagram leads to the expression (after disorder average):
\begin{align}
\tilde{\Sigma}(p)=& - \lambda_{z} \, n_{i} \, v_{0}^{2} \, \sum_{\vec{p}'}  \,  i \, \left[  \Gamma_{\mathrm{M}}(\vec{p},\vec{p}'),G(p')\right] \nonumber \\
&+ (-e\vec{E})\partial_{\epsilon} \cdot  \frac{1}{2} \left\{ \boldsymbol{\sigma}\times(\vec{p}-\vec{p}'),\tilde{G}(p')\right\}  
\end{align}
The resulting collision integral is given by the sum $I_{\mathrm{SJ}}[n_{\vec{p}}]=I_{\mathrm{SJ}}^{0}[n_{\vec{p}}]+I_{\mathrm{SJ}}^{1}[n_{\vec{p}}]$, where
\begin{align}
I_{\mathrm{SJ}}^{1}[n_{\vec{p}}]&= \lambda_{z} \, \pi \, n_{i} \, v_{0}^{2} \, \sum_{\vec{p}'} \partial_{\epsilon_p} \delta(\epsilon_p -\epsilon_p')  \nonumber \\
&\times \, \left\{ ( -e\vec{E} )\cdot \left( \boldsymbol{\sigma}\times( \vec{p}-\vec{p}') \right), n_{\vec{p}}-n_{\vec{p}'}\right\} \\ \nonumber
I_{\mathrm{SJ}}^{0}[n_{\vec{p}}] &= 2\pi  n_{i } \, \lambda_{z} \, v_{0}^2 \, i \, \sum_{\vec{p}'}  \delta(\epsilon_p -\epsilon_p')  [n_{\vec{p}'} , \Gamma_{\mathrm{M}}(\vec{p},\vec{p}')] 
\end{align}
\subsection{Skew scattering: Mott-scalar-scalar  self-energy}
The contribution to self-energy that gives rise to Mott's skew scattering  is given by the following expression:
\begin{align}
\tilde{\Sigma}(p)&=-i \, \lambda_z \, n_s \, v_{0}^3 \, \sum_{\vec{p}_a  \vec{p}_b}\bigg( 
\tilde{G}(p_a) \, \Gamma_{\mathrm{M}}(\vec{p}_a , \vec{p}_b) \, \tilde{G}(p_b) \nonumber \\
+& \tilde{G}(p_a) \, \tilde{G}(p_b) \, \Gamma_{\mathrm{M}}(\vec{p}_b , \vec{p}) 
+\Gamma_{\mathrm{M}}(\vec{p} , \vec{p}_a) \tilde{G}(p_a)  \tilde{G}(p_b)\bigg). \nonumber
\end{align}
The resulting collision integral is given by
\begin{equation}
I_{\mathrm{SK}}[n_{\vec{p}}]=-2\pi^2 \, n_s \, v_{0}^3 \, N_{0} \, \lambda_{z} \, \sum_{p'} \delta(\epsilon_p -\epsilon_p')
\left\{ \Gamma_{\mathrm{M}}(\vec{p}',\vec{p}) , n_{\vec{p}'} \right\},
\end{equation}
Collecting all the contributions to collision integral from the seven self-energy diagrams, we obtain the  complete collision integral used in the main text of the article.

\appendix

\bibliography{reference.bib}

\begin{thebibliography}{56}%
\makeatletter
\providecommand \@ifxundefined [1]{%
 \@ifx{#1\undefined}
}%
\providecommand \@ifnum [1]{%
 \ifnum #1\expandafter \@firstoftwo
 \else \expandafter \@secondoftwo
 \fi
}%
\providecommand \@ifx [1]{%
 \ifx #1\expandafter \@firstoftwo
 \else \expandafter \@secondoftwo
 \fi
}%
\providecommand \natexlab [1]{#1}%
\providecommand \enquote  [1]{``#1''}%
\providecommand \bibnamefont  [1]{#1}%
\providecommand \bibfnamefont [1]{#1}%
\providecommand \citenamefont [1]{#1}%
\providecommand \href@noop [0]{\@secondoftwo}%
\providecommand \href [0]{\begingroup \@sanitize@url \@href}%
\providecommand \@href[1]{\@@startlink{#1}\@@href}%
\providecommand \@@href[1]{\endgroup#1\@@endlink}%
\providecommand \@sanitize@url [0]{\catcode `\\12\catcode `\$12\catcode
  `\&12\catcode `\#12\catcode `\^12\catcode `\_12\catcode `\%12\relax}%
\providecommand \@@startlink[1]{}%
\providecommand \@@endlink[0]{}%
\providecommand \url  [0]{\begingroup\@sanitize@url \@url }%
\providecommand \@url [1]{\endgroup\@href {#1}{\urlprefix }}%
\providecommand \urlprefix  [0]{URL }%
\providecommand \Eprint [0]{\href }%
\providecommand \doibase [0]{http://dx.doi.org/}%
\providecommand \selectlanguage [0]{\@gobble}%
\providecommand \bibinfo  [0]{\@secondoftwo}%
\providecommand \bibfield  [0]{\@secondoftwo}%
\providecommand \translation [1]{[#1]}%
\providecommand \BibitemOpen [0]{}%
\providecommand \bibitemStop [0]{}%
\providecommand \bibitemNoStop [0]{.\EOS\space}%
\providecommand \EOS [0]{\spacefactor3000\relax}%
\providecommand \BibitemShut  [1]{\csname bibitem#1\endcsname}%
\let\auto@bib@innerbib\@empty
\bibitem [{\citenamefont {Soumyanarayanan}\ \emph {et~al.}(2016)\citenamefont
  {Soumyanarayanan}, \citenamefont {Reyren}, \citenamefont {Fert},\ and\
  \citenamefont {Panagopoulos}}]{soumyanarayanan2016emergent}%
  \BibitemOpen
  \bibfield  {author} {\bibinfo {author} {\bibfnamefont {A.}~\bibnamefont
  {Soumyanarayanan}}, \bibinfo {author} {\bibfnamefont {N.}~\bibnamefont
  {Reyren}}, \bibinfo {author} {\bibfnamefont {A.}~\bibnamefont {Fert}}, \ and\
  \bibinfo {author} {\bibfnamefont {C.}~\bibnamefont {Panagopoulos}},\
  }\href@noop {} {\bibfield  {journal} {\bibinfo  {journal} {Nature}\ }\textbf
  {\bibinfo {volume} {539}},\ \bibinfo {pages} {509} (\bibinfo {year}
  {2016})}\BibitemShut {NoStop}%
\bibitem [{\citenamefont {Wunderlich}\ \emph {et~al.}(2010)\citenamefont
  {Wunderlich}, \citenamefont {Park}, \citenamefont {Irvine}, \citenamefont
  {Z{\^a}rbo}, \citenamefont {Rozkotov{\'a}}, \citenamefont {Nemec},
  \citenamefont {Nov{\'a}k}, \citenamefont {Sinova},\ and\ \citenamefont
  {Jungwirth}}]{wunderlich2010spin}%
  \BibitemOpen
  \bibfield  {author} {\bibinfo {author} {\bibfnamefont {J.}~\bibnamefont
  {Wunderlich}}, \bibinfo {author} {\bibfnamefont {B.-G.}\ \bibnamefont
  {Park}}, \bibinfo {author} {\bibfnamefont {A.~C.}\ \bibnamefont {Irvine}},
  \bibinfo {author} {\bibfnamefont {L.~P.}\ \bibnamefont {Z{\^a}rbo}}, \bibinfo
  {author} {\bibfnamefont {E.}~\bibnamefont {Rozkotov{\'a}}}, \bibinfo {author}
  {\bibfnamefont {P.}~\bibnamefont {Nemec}}, \bibinfo {author} {\bibfnamefont
  {V.}~\bibnamefont {Nov{\'a}k}}, \bibinfo {author} {\bibfnamefont
  {J.}~\bibnamefont {Sinova}}, \ and\ \bibinfo {author} {\bibfnamefont
  {T.}~\bibnamefont {Jungwirth}},\ }\href@noop {} {\bibfield  {journal}
  {\bibinfo  {journal} {Science}\ }\textbf {\bibinfo {volume} {330}},\ \bibinfo
  {pages} {1801} (\bibinfo {year} {2010})}\BibitemShut {NoStop}%
\bibitem [{\citenamefont {Sinova}\ \emph {et~al.}(2015)\citenamefont {Sinova},
  \citenamefont {Valenzuela}, \citenamefont {Wunderlich}, \citenamefont
  {Back},\ and\ \citenamefont {Jungwirth}}]{sinova2015spin}%
  \BibitemOpen
  \bibfield  {author} {\bibinfo {author} {\bibfnamefont {J.}~\bibnamefont
  {Sinova}}, \bibinfo {author} {\bibfnamefont {S.~O.}\ \bibnamefont
  {Valenzuela}}, \bibinfo {author} {\bibfnamefont {J.}~\bibnamefont
  {Wunderlich}}, \bibinfo {author} {\bibfnamefont {C.~H.}\ \bibnamefont
  {Back}}, \ and\ \bibinfo {author} {\bibfnamefont {T.}~\bibnamefont
  {Jungwirth}},\ }\href {\doibase 10.1103/RevModPhys.87.1213} {\bibfield
  {journal} {\bibinfo  {journal} {Rev. Mod. Phys.}\ }\textbf {\bibinfo {volume}
  {87}},\ \bibinfo {pages} {1213} (\bibinfo {year} {2015})}\BibitemShut
  {NoStop}%
\bibitem [{\citenamefont {Bychkov}\ and\ \citenamefont
  {Rashba}(1984)}]{rashba1984}%
  \BibitemOpen
  \bibfield  {author} {\bibinfo {author} {\bibfnamefont {Y.}~\bibnamefont
  {Bychkov}}\ and\ \bibinfo {author} {\bibfnamefont {E.}~\bibnamefont
  {Rashba}},\ }\href@noop {} {\bibfield  {journal} {\bibinfo  {journal} {Sov.
  Phys. JETP}\ }\textbf {\bibinfo {volume} {39}},\ \bibinfo {pages} {78}
  (\bibinfo {year} {1984})}\BibitemShut {NoStop}%
\bibitem [{\citenamefont {Edelstein}(1990)}]{Edelstein1990233}%
  \BibitemOpen
  \bibfield  {author} {\bibinfo {author} {\bibfnamefont {V.}~\bibnamefont
  {Edelstein}},\ }\href {\doibase
  http://dx.doi.org/10.1016/0038-1098(90)90963-C} {\bibfield  {journal}
  {\bibinfo  {journal} {Solid State Communications}\ }\textbf {\bibinfo
  {volume} {73}},\ \bibinfo {pages} {233 } (\bibinfo {year}
  {1990})}\BibitemShut {NoStop}%
\bibitem [{\citenamefont {Sinova}\ \emph {et~al.}(2004)\citenamefont {Sinova},
  \citenamefont {Culcer}, \citenamefont {Niu}, \citenamefont {Sinitsyn},
  \citenamefont {Jungwirth},\ and\ \citenamefont
  {MacDonald}}]{sinova2004universal}%
  \BibitemOpen
  \bibfield  {author} {\bibinfo {author} {\bibfnamefont {J.}~\bibnamefont
  {Sinova}}, \bibinfo {author} {\bibfnamefont {D.}~\bibnamefont {Culcer}},
  \bibinfo {author} {\bibfnamefont {Q.}~\bibnamefont {Niu}}, \bibinfo {author}
  {\bibfnamefont {N.}~\bibnamefont {Sinitsyn}}, \bibinfo {author}
  {\bibfnamefont {T.}~\bibnamefont {Jungwirth}}, \ and\ \bibinfo {author}
  {\bibfnamefont {A.}~\bibnamefont {MacDonald}},\ }\href@noop {} {\bibfield
  {journal} {\bibinfo  {journal} {Phys. Rev. Lett.}\ }\textbf {\bibinfo
  {volume} {92}},\ \bibinfo {pages} {126603} (\bibinfo {year}
  {2004})}\BibitemShut {NoStop}%
\bibitem [{\citenamefont {Hirsch}(1999)}]{hirsch1999spin}%
  \BibitemOpen
  \bibfield  {author} {\bibinfo {author} {\bibfnamefont {J.}~\bibnamefont
  {Hirsch}},\ }\href@noop {} {\bibfield  {journal} {\bibinfo  {journal} {Phys.
  Rev. Lett.}\ }\textbf {\bibinfo {volume} {83}},\ \bibinfo {pages} {1834}
  (\bibinfo {year} {1999})}\BibitemShut {NoStop}%
\bibitem [{\citenamefont {Nagaosa}\ \emph {et~al.}(2010)\citenamefont
  {Nagaosa}, \citenamefont {Sinova}, \citenamefont {Onoda}, \citenamefont
  {MacDonald},\ and\ \citenamefont {Ong}}]{Nagaosa_RevModPhys.82.1539}%
  \BibitemOpen
  \bibfield  {author} {\bibinfo {author} {\bibfnamefont {N.}~\bibnamefont
  {Nagaosa}}, \bibinfo {author} {\bibfnamefont {J.}~\bibnamefont {Sinova}},
  \bibinfo {author} {\bibfnamefont {S.}~\bibnamefont {Onoda}}, \bibinfo
  {author} {\bibfnamefont {A.~H.}\ \bibnamefont {MacDonald}}, \ and\ \bibinfo
  {author} {\bibfnamefont {N.~P.}\ \bibnamefont {Ong}},\ }\href {\doibase
  10.1103/RevModPhys.82.1539} {\bibfield  {journal} {\bibinfo  {journal} {Rev.
  Mod. Phys.}\ }\textbf {\bibinfo {volume} {82}},\ \bibinfo {pages} {1539}
  (\bibinfo {year} {2010})}\BibitemShut {NoStop}%
\bibitem [{\citenamefont {Zhang}\ and\ \citenamefont
  {Fert}(2016)}]{fert_zhang_2016}%
  \BibitemOpen
  \bibfield  {author} {\bibinfo {author} {\bibfnamefont {S.}~\bibnamefont
  {Zhang}}\ and\ \bibinfo {author} {\bibfnamefont {A.}~\bibnamefont {Fert}},\
  }\href {\doibase 10.1103/PhysRevB.94.184423} {\bibfield  {journal} {\bibinfo
  {journal} {Phys. Rev. B}\ }\textbf {\bibinfo {volume} {94}},\ \bibinfo
  {pages} {184423} (\bibinfo {year} {2016})}\BibitemShut {NoStop}%
\bibitem [{\citenamefont {Maleki}\ \emph {et~al.}(2016)\citenamefont {Maleki},
  \citenamefont {Raimondi},\ and\ \citenamefont {Shen}}]{amin2016the}%
  \BibitemOpen
  \bibfield  {author} {\bibinfo {author} {\bibfnamefont {A.}~\bibnamefont
  {Maleki}}, \bibinfo {author} {\bibfnamefont {R.}~\bibnamefont {Raimondi}}, \
  and\ \bibinfo {author} {\bibfnamefont {K.}~\bibnamefont {Shen}},\ }\href@noop
  {} {\bibfield  {journal} {\bibinfo  {journal} {arXiv:1610.08258v1}\ }
  (\bibinfo {year} {2016})}\BibitemShut {NoStop}%
\bibitem [{\citenamefont {Song}\ \emph {et~al.}(2016)\citenamefont {Song},
  \citenamefont {Mi}, \citenamefont {Zhao}, \citenamefont {Su}, \citenamefont
  {Yuan}, \citenamefont {Xing}, \citenamefont {Chen}, \citenamefont {Wang},
  \citenamefont {Wu}, \citenamefont {Chen} \emph {et~al.}}]{song2016spin}%
  \BibitemOpen
  \bibfield  {author} {\bibinfo {author} {\bibfnamefont {Q.}~\bibnamefont
  {Song}}, \bibinfo {author} {\bibfnamefont {J.}~\bibnamefont {Mi}}, \bibinfo
  {author} {\bibfnamefont {D.}~\bibnamefont {Zhao}}, \bibinfo {author}
  {\bibfnamefont {T.}~\bibnamefont {Su}}, \bibinfo {author} {\bibfnamefont
  {W.}~\bibnamefont {Yuan}}, \bibinfo {author} {\bibfnamefont {W.}~\bibnamefont
  {Xing}}, \bibinfo {author} {\bibfnamefont {Y.}~\bibnamefont {Chen}}, \bibinfo
  {author} {\bibfnamefont {T.}~\bibnamefont {Wang}}, \bibinfo {author}
  {\bibfnamefont {T.}~\bibnamefont {Wu}}, \bibinfo {author} {\bibfnamefont
  {X.~H.}\ \bibnamefont {Chen}},  \emph {et~al.},\ }\href@noop {} {\bibfield
  {journal} {\bibinfo  {journal} {Nature Communications}\ }\textbf {\bibinfo
  {volume} {7}} (\bibinfo {year} {2016})}\BibitemShut {NoStop}%
\bibitem [{\citenamefont {Song}\ \emph {et~al.}(2017)\citenamefont {Song},
  \citenamefont {Zhang}, \citenamefont {Su}, \citenamefont {Yuan},
  \citenamefont {Chen}, \citenamefont {Xing}, \citenamefont {Shi},
  \citenamefont {Sun},\ and\ \citenamefont {Han}}]{song2017observation}%
  \BibitemOpen
  \bibfield  {author} {\bibinfo {author} {\bibfnamefont {Q.}~\bibnamefont
  {Song}}, \bibinfo {author} {\bibfnamefont {H.}~\bibnamefont {Zhang}},
  \bibinfo {author} {\bibfnamefont {T.}~\bibnamefont {Su}}, \bibinfo {author}
  {\bibfnamefont {W.}~\bibnamefont {Yuan}}, \bibinfo {author} {\bibfnamefont
  {Y.}~\bibnamefont {Chen}}, \bibinfo {author} {\bibfnamefont {W.}~\bibnamefont
  {Xing}}, \bibinfo {author} {\bibfnamefont {J.}~\bibnamefont {Shi}}, \bibinfo
  {author} {\bibfnamefont {J.}~\bibnamefont {Sun}}, \ and\ \bibinfo {author}
  {\bibfnamefont {W.}~\bibnamefont {Han}},\ }\href@noop {} {\bibfield
  {journal} {\bibinfo  {journal} {Science Advances}\ }\textbf {\bibinfo
  {volume} {3}},\ \bibinfo {pages} {e1602312} (\bibinfo {year}
  {2017})}\BibitemShut {NoStop}%
\bibitem [{\citenamefont {Gorini}\ \emph {et~al.}(2017)\citenamefont {Gorini},
  \citenamefont {Maleki}, \citenamefont {Shen}, \citenamefont {Tokatly},
  \citenamefont {Vignale},\ and\ \citenamefont {Raimondi}}]{cosimo2017theory}%
  \BibitemOpen
  \bibfield  {author} {\bibinfo {author} {\bibfnamefont {C.}~\bibnamefont
  {Gorini}}, \bibinfo {author} {\bibfnamefont {A.}~\bibnamefont {Maleki}},
  \bibinfo {author} {\bibfnamefont {K.}~\bibnamefont {Shen}}, \bibinfo {author}
  {\bibfnamefont {I.}~\bibnamefont {Tokatly}}, \bibinfo {author} {\bibfnamefont
  {G.}~\bibnamefont {Vignale}}, \ and\ \bibinfo {author} {\bibfnamefont
  {R.}~\bibnamefont {Raimondi}},\ }\href@noop {} {\bibfield  {journal}
  {\bibinfo  {journal} {arXiv:1702.04887v1}\ } (\bibinfo {year}
  {2017})}\BibitemShut {NoStop}%
\bibitem [{\citenamefont {Mishchenko}\ \emph {et~al.}(2004)\citenamefont
  {Mishchenko}, \citenamefont {Shytov},\ and\ \citenamefont
  {Halperin}}]{PhysRevLett.93.226602}%
  \BibitemOpen
  \bibfield  {author} {\bibinfo {author} {\bibfnamefont {E.~G.}\ \bibnamefont
  {Mishchenko}}, \bibinfo {author} {\bibfnamefont {A.~V.}\ \bibnamefont
  {Shytov}}, \ and\ \bibinfo {author} {\bibfnamefont {B.~I.}\ \bibnamefont
  {Halperin}},\ }\href {\doibase 10.1103/PhysRevLett.93.226602} {\bibfield
  {journal} {\bibinfo  {journal} {Phys. Rev. Lett.}\ }\textbf {\bibinfo
  {volume} {93}},\ \bibinfo {pages} {226602} (\bibinfo {year}
  {2004})}\BibitemShut {NoStop}%
\bibitem [{\citenamefont {Ganichev}\ \emph {et~al.}(2016)\citenamefont
  {Ganichev}, \citenamefont {Trushin},\ and\ \citenamefont
  {Schliemann}}]{ganichev2016spin}%
  \BibitemOpen
  \bibfield  {author} {\bibinfo {author} {\bibfnamefont {S.~D.}\ \bibnamefont
  {Ganichev}}, \bibinfo {author} {\bibfnamefont {M.}~\bibnamefont {Trushin}}, \
  and\ \bibinfo {author} {\bibfnamefont {J.}~\bibnamefont {Schliemann}},\
  }\href@noop {} {\bibfield  {journal} {\bibinfo  {journal} {arXiv:1606.02043}\
  } (\bibinfo {year} {2016})}\BibitemShut {NoStop}%
\bibitem [{\citenamefont {Manuel~Offidani}(2017)}]{manuel2017}%
  \BibitemOpen
  \bibfield  {author} {\bibinfo {author} {\bibfnamefont {R.~R. A.~F.}\
  \bibnamefont {Manuel~Offidani}, \bibfnamefont {Mirco~Milletarí}},\
  }\href@noop {} {\bibfield  {journal} {\bibinfo  {journal} {arXiv:1706.08973}\
  } (\bibinfo {year} {2017})}\BibitemShut {NoStop}%
\bibitem [{\citenamefont {Castro~Neto}\ and\ \citenamefont
  {Guinea}(2009)}]{netoguinea09}%
  \BibitemOpen
  \bibfield  {author} {\bibinfo {author} {\bibfnamefont {A.~H.}\ \bibnamefont
  {Castro~Neto}}\ and\ \bibinfo {author} {\bibfnamefont {F.}~\bibnamefont
  {Guinea}},\ }\href {\doibase 10.1103/PhysRevLett.103.026804} {\bibfield
  {journal} {\bibinfo  {journal} {Phys. Rev. Lett.}\ }\textbf {\bibinfo
  {volume} {103}},\ \bibinfo {pages} {026804} (\bibinfo {year}
  {2009})}\BibitemShut {NoStop}%
\bibitem [{\citenamefont {Weeks}\ \emph {et~al.}(2011)\citenamefont {Weeks},
  \citenamefont {Hu}, \citenamefont {Alicea}, \citenamefont {Franz},\ and\
  \citenamefont {Wu}}]{weeks2011engineering}%
  \BibitemOpen
  \bibfield  {author} {\bibinfo {author} {\bibfnamefont {C.}~\bibnamefont
  {Weeks}}, \bibinfo {author} {\bibfnamefont {J.}~\bibnamefont {Hu}}, \bibinfo
  {author} {\bibfnamefont {J.}~\bibnamefont {Alicea}}, \bibinfo {author}
  {\bibfnamefont {M.}~\bibnamefont {Franz}}, \ and\ \bibinfo {author}
  {\bibfnamefont {R.}~\bibnamefont {Wu}},\ }\href@noop {} {\bibfield  {journal}
  {\bibinfo  {journal} {Phys. Rev. X}\ }\textbf {\bibinfo {volume} {1}},\
  \bibinfo {pages} {021001} (\bibinfo {year} {2011})}\BibitemShut {NoStop}%
\bibitem [{\citenamefont {Ferreira}\ \emph {et~al.}(2014)\citenamefont
  {Ferreira}, \citenamefont {Rappoport}, \citenamefont {Cazalilla},\ and\
  \citenamefont {Neto}}]{ferreira2014extrinsic}%
  \BibitemOpen
  \bibfield  {author} {\bibinfo {author} {\bibfnamefont {A.}~\bibnamefont
  {Ferreira}}, \bibinfo {author} {\bibfnamefont {T.~G.}\ \bibnamefont
  {Rappoport}}, \bibinfo {author} {\bibfnamefont {M.~A.}\ \bibnamefont
  {Cazalilla}}, \ and\ \bibinfo {author} {\bibfnamefont {A.~C.}\ \bibnamefont
  {Neto}},\ }\href@noop {} {\bibfield  {journal} {\bibinfo  {journal} {Phys.
  Rev. Lett.}\ }\textbf {\bibinfo {volume} {112}},\ \bibinfo {pages} {066601}
  (\bibinfo {year} {2014})}\BibitemShut {NoStop}%
\bibitem [{\citenamefont {Huang}\ \emph
  {et~al.}(2016{\natexlab{a}})\citenamefont {Huang}, \citenamefont {Chong},\
  and\ \citenamefont {Cazalilla}}]{chunli2016direct}%
  \BibitemOpen
  \bibfield  {author} {\bibinfo {author} {\bibfnamefont {C.}~\bibnamefont
  {Huang}}, \bibinfo {author} {\bibfnamefont {Y.~D.}\ \bibnamefont {Chong}}, \
  and\ \bibinfo {author} {\bibfnamefont {M.~A.}\ \bibnamefont {Cazalilla}},\
  }\href {\doibase 10.1103/PhysRevB.94.085414} {\bibfield  {journal} {\bibinfo
  {journal} {Phys. Rev. B}\ }\textbf {\bibinfo {volume} {94}},\ \bibinfo
  {pages} {085414} (\bibinfo {year} {2016}{\natexlab{a}})}\BibitemShut
  {NoStop}%
\bibitem [{\citenamefont {Milletar\`{\i}}\ and\ \citenamefont
  {Ferreira}(2016{\natexlab{a}})}]{mirco2016crossover}%
  \BibitemOpen
  \bibfield  {author} {\bibinfo {author} {\bibfnamefont {M.}~\bibnamefont
  {Milletar\`{\i}}}\ and\ \bibinfo {author} {\bibfnamefont {A.}~\bibnamefont
  {Ferreira}},\ }\href {\doibase 10.1103/PhysRevB.94.201402} {\bibfield
  {journal} {\bibinfo  {journal} {Phys. Rev. B}\ }\textbf {\bibinfo {volume}
  {94}},\ \bibinfo {pages} {201402} (\bibinfo {year}
  {2016}{\natexlab{a}})}\BibitemShut {NoStop}%
\bibitem [{\citenamefont {Milletar\`{\i}}\ and\ \citenamefont
  {Ferreira}(2016{\natexlab{b}})}]{mirco2016quantum}%
  \BibitemOpen
  \bibfield  {author} {\bibinfo {author} {\bibfnamefont {M.}~\bibnamefont
  {Milletar\`{\i}}}\ and\ \bibinfo {author} {\bibfnamefont {A.}~\bibnamefont
  {Ferreira}},\ }\href@noop {} {\bibfield  {journal} {\bibinfo  {journal}
  {Phys. Rev. B}\ }\textbf {\bibinfo {volume} {94}},\ \bibinfo {pages} {134202}
  (\bibinfo {year} {2016}{\natexlab{b}})}\BibitemShut {NoStop}%
\bibitem [{\citenamefont {Huang}\ \emph {et~al.}(2017)\citenamefont {Huang},
  \citenamefont {Chong},\ and\ \citenamefont
  {Cazalilla}}]{chunli2017anomalous}%
  \BibitemOpen
  \bibfield  {author} {\bibinfo {author} {\bibfnamefont {C.}~\bibnamefont
  {Huang}}, \bibinfo {author} {\bibfnamefont {Y.}~\bibnamefont {Chong}}, \ and\
  \bibinfo {author} {\bibfnamefont {M.}~\bibnamefont {Cazalilla}},\ }\href@noop
  {} {\bibfield  {journal} {\bibinfo  {journal} {arXiv:1702.04955}\ } (\bibinfo
  {year} {2017})}\BibitemShut {NoStop}%
\bibitem [{\citenamefont {Bergeret}\ and\ \citenamefont
  {Tokatly}(2016)}]{PhysRevB.94.180502}%
  \BibitemOpen
  \bibfield  {author} {\bibinfo {author} {\bibfnamefont {F.~S.}\ \bibnamefont
  {Bergeret}}\ and\ \bibinfo {author} {\bibfnamefont {I.~V.}\ \bibnamefont
  {Tokatly}},\ }\href {\doibase 10.1103/PhysRevB.94.180502} {\bibfield
  {journal} {\bibinfo  {journal} {Phys. Rev. B}\ }\textbf {\bibinfo {volume}
  {94}},\ \bibinfo {pages} {180502} (\bibinfo {year} {2016})}\BibitemShut
  {NoStop}%
\bibitem [{\citenamefont {Ilya~Tokatly}(2017)}]{tokatly2017usadel}%
  \BibitemOpen
  \bibfield  {author} {\bibinfo {author} {\bibfnamefont {a.}~\bibnamefont
  {Ilya~Tokatly}},\ }\href@noop {} {} (\bibinfo {year} {2017})\BibitemShut
  {NoStop}%
\bibitem [{\citenamefont {Zyuzin}\ \emph {et~al.}(2016)\citenamefont {Zyuzin},
  \citenamefont {Alidoust},\ and\ \citenamefont {Loss}}]{PhysRevB.93.214502}%
  \BibitemOpen
  \bibfield  {author} {\bibinfo {author} {\bibfnamefont {A.}~\bibnamefont
  {Zyuzin}}, \bibinfo {author} {\bibfnamefont {M.}~\bibnamefont {Alidoust}}, \
  and\ \bibinfo {author} {\bibfnamefont {D.}~\bibnamefont {Loss}},\ }\href
  {\doibase 10.1103/PhysRevB.93.214502} {\bibfield  {journal} {\bibinfo
  {journal} {Phys. Rev. B}\ }\textbf {\bibinfo {volume} {93}},\ \bibinfo
  {pages} {214502} (\bibinfo {year} {2016})}\BibitemShut {NoStop}%
\bibitem [{\citenamefont {Bobkova}\ \emph {et~al.}(2016)\citenamefont
  {Bobkova}, \citenamefont {Bobkov}, \citenamefont {Zyuzin},\ and\
  \citenamefont {Alidoust}}]{PhysRevB.94.134506}%
  \BibitemOpen
  \bibfield  {author} {\bibinfo {author} {\bibfnamefont {I.~V.}\ \bibnamefont
  {Bobkova}}, \bibinfo {author} {\bibfnamefont {A.~M.}\ \bibnamefont {Bobkov}},
  \bibinfo {author} {\bibfnamefont {A.~A.}\ \bibnamefont {Zyuzin}}, \ and\
  \bibinfo {author} {\bibfnamefont {M.}~\bibnamefont {Alidoust}},\ }\href
  {\doibase 10.1103/PhysRevB.94.134506} {\bibfield  {journal} {\bibinfo
  {journal} {Phys. Rev. B}\ }\textbf {\bibinfo {volume} {94}},\ \bibinfo
  {pages} {134506} (\bibinfo {year} {2016})}\BibitemShut {NoStop}%
\bibitem [{\citenamefont {Balakrishnan}\ \emph {et~al.}(2013)\citenamefont
  {Balakrishnan}, \citenamefont {Koon}, \citenamefont {Jaiswal}, \citenamefont
  {Neto},\ and\ \citenamefont {\"Ozyilmaz}}]{balakrishnan_colossal}%
  \BibitemOpen
  \bibfield  {author} {\bibinfo {author} {\bibfnamefont {J.}~\bibnamefont
  {Balakrishnan}}, \bibinfo {author} {\bibfnamefont {G.~K.~W.}\ \bibnamefont
  {Koon}}, \bibinfo {author} {\bibfnamefont {M.}~\bibnamefont {Jaiswal}},
  \bibinfo {author} {\bibfnamefont {A.~H.~C.}\ \bibnamefont {Neto}}, \ and\
  \bibinfo {author} {\bibfnamefont {B.}~\bibnamefont {\"Ozyilmaz}},\
  }\href@noop {} {\bibfield  {journal} {\bibinfo  {journal} {Nature Physics}\
  }\textbf {\bibinfo {volume} {9}},\ \bibinfo {pages} {284} (\bibinfo {year}
  {2013})}\BibitemShut {NoStop}%
\bibitem [{\citenamefont {Glazov}\ \emph {et~al.}(2010)\citenamefont {Glazov},
  \citenamefont {Sherman},\ and\ \citenamefont {Dugaev}}]{glazov2010two}%
  \BibitemOpen
  \bibfield  {author} {\bibinfo {author} {\bibfnamefont {M.}~\bibnamefont
  {Glazov}}, \bibinfo {author} {\bibfnamefont {E.~Y.}\ \bibnamefont {Sherman}},
  \ and\ \bibinfo {author} {\bibfnamefont {V.}~\bibnamefont {Dugaev}},\
  }\href@noop {} {\bibfield  {journal} {\bibinfo  {journal} {Physica E:
  Low-dimensional Systems and Nanostructures}\ }\textbf {\bibinfo {volume}
  {42}},\ \bibinfo {pages} {2157} (\bibinfo {year} {2010})}\BibitemShut
  {NoStop}%
\bibitem [{\citenamefont {Wang}\ \emph {et~al.}(2016)\citenamefont {Wang},
  \citenamefont {Ki}, \citenamefont {Khoo}, \citenamefont {Mauro},
  \citenamefont {Berger}, \citenamefont {Levitov},\ and\ \citenamefont
  {Morpurgo}}]{wang2016origin}%
  \BibitemOpen
  \bibfield  {author} {\bibinfo {author} {\bibfnamefont {Z.}~\bibnamefont
  {Wang}}, \bibinfo {author} {\bibfnamefont {D.-K.}\ \bibnamefont {Ki}},
  \bibinfo {author} {\bibfnamefont {J.~Y.}\ \bibnamefont {Khoo}}, \bibinfo
  {author} {\bibfnamefont {D.}~\bibnamefont {Mauro}}, \bibinfo {author}
  {\bibfnamefont {H.}~\bibnamefont {Berger}}, \bibinfo {author} {\bibfnamefont
  {L.~S.}\ \bibnamefont {Levitov}}, \ and\ \bibinfo {author} {\bibfnamefont
  {A.~F.}\ \bibnamefont {Morpurgo}},\ }\href {\doibase
  10.1103/PhysRevX.6.041020} {\bibfield  {journal} {\bibinfo  {journal} {Phys.
  Rev. X}\ }\textbf {\bibinfo {volume} {6}},\ \bibinfo {pages} {041020}
  (\bibinfo {year} {2016})}\BibitemShut {NoStop}%
\bibitem [{\citenamefont {Yang}\ \emph
  {et~al.}(2016{\natexlab{a}})\citenamefont {Yang}, \citenamefont {Tu},
  \citenamefont {Kim}, \citenamefont {Wu}, \citenamefont {Wang}, \citenamefont
  {Alicea}, \citenamefont {Bockrath},\ and\ \citenamefont
  {Shi}}]{Yang_G_on_TMD2016}%
  \BibitemOpen
  \bibfield  {author} {\bibinfo {author} {\bibfnamefont {B.}~\bibnamefont
  {Yang}}, \bibinfo {author} {\bibfnamefont {M.-F.}\ \bibnamefont {Tu}},
  \bibinfo {author} {\bibfnamefont {J.}~\bibnamefont {Kim}}, \bibinfo {author}
  {\bibfnamefont {Y.}~\bibnamefont {Wu}}, \bibinfo {author} {\bibfnamefont
  {H.}~\bibnamefont {Wang}}, \bibinfo {author} {\bibfnamefont {J.}~\bibnamefont
  {Alicea}}, \bibinfo {author} {\bibfnamefont {M.}~\bibnamefont {Bockrath}}, \
  and\ \bibinfo {author} {\bibfnamefont {J.}~\bibnamefont {Shi}},\ }\href@noop
  {} {\bibfield  {journal} {\bibinfo  {journal} {2D Mater.}\ }\textbf {\bibinfo
  {volume} {3}},\ \bibinfo {pages} {031012} (\bibinfo {year}
  {2016}{\natexlab{a}})}\BibitemShut {NoStop}%
\bibitem [{\citenamefont {Castro~Neto}\ \emph {et~al.}(2009)\citenamefont
  {Castro~Neto}, \citenamefont {Guinea}, \citenamefont {Peres}, \citenamefont
  {Novoselov},\ and\ \citenamefont {Geim}}]{neto2009electronic}%
  \BibitemOpen
  \bibfield  {author} {\bibinfo {author} {\bibfnamefont {A.~H.}\ \bibnamefont
  {Castro~Neto}}, \bibinfo {author} {\bibfnamefont {F.}~\bibnamefont {Guinea}},
  \bibinfo {author} {\bibfnamefont {N.}~\bibnamefont {Peres}}, \bibinfo
  {author} {\bibfnamefont {K.~S.}\ \bibnamefont {Novoselov}}, \ and\ \bibinfo
  {author} {\bibfnamefont {A.~K.}\ \bibnamefont {Geim}},\ }\href@noop {}
  {\bibfield  {journal} {\bibinfo  {journal} {Rev. Mod. Phys.}\ }\textbf
  {\bibinfo {volume} {81}},\ \bibinfo {pages} {109} (\bibinfo {year}
  {2009})}\BibitemShut {NoStop}%
\bibitem [{\citenamefont {Das~Sarma}\ \emph {et~al.}(2011)\citenamefont
  {Das~Sarma}, \citenamefont {Adam}, \citenamefont {Hwang},\ and\ \citenamefont
  {Rossi}}]{das2011electronic}%
  \BibitemOpen
  \bibfield  {author} {\bibinfo {author} {\bibfnamefont {S.}~\bibnamefont
  {Das~Sarma}}, \bibinfo {author} {\bibfnamefont {S.}~\bibnamefont {Adam}},
  \bibinfo {author} {\bibfnamefont {E.~H.}\ \bibnamefont {Hwang}}, \ and\
  \bibinfo {author} {\bibfnamefont {E.}~\bibnamefont {Rossi}},\ }\href
  {\doibase 10.1103/RevModPhys.83.407} {\bibfield  {journal} {\bibinfo
  {journal} {Rev. Mod. Phys.}\ }\textbf {\bibinfo {volume} {83}},\ \bibinfo
  {pages} {407} (\bibinfo {year} {2011})}\BibitemShut {NoStop}%
\bibitem [{\citenamefont {Tsui}\ \emph {et~al.}(1982)\citenamefont {Tsui},
  \citenamefont {Stormer},\ and\ \citenamefont {Gossard}}]{tsui1982FQH}%
  \BibitemOpen
  \bibfield  {author} {\bibinfo {author} {\bibfnamefont {D.~C.}\ \bibnamefont
  {Tsui}}, \bibinfo {author} {\bibfnamefont {H.~L.}\ \bibnamefont {Stormer}}, \
  and\ \bibinfo {author} {\bibfnamefont {A.~C.}\ \bibnamefont {Gossard}},\
  }\href {\doibase 10.1103/PhysRevLett.48.1559} {\bibfield  {journal} {\bibinfo
   {journal} {Phys. Rev. Lett.}\ }\textbf {\bibinfo {volume} {48}},\ \bibinfo
  {pages} {1559} (\bibinfo {year} {1982})}\BibitemShut {NoStop}%
\bibitem [{\citenamefont {Sherman}(2003)}]{sherman2002}%
  \BibitemOpen
  \bibfield  {author} {\bibinfo {author} {\bibfnamefont {E.~Y.}\ \bibnamefont
  {Sherman}},\ }\href {\doibase http://dx.doi.org/10.1063/1.1533839} {\bibfield
   {journal} {\bibinfo  {journal} {Applied Physics Letters}\ }\textbf {\bibinfo
  {volume} {82}},\ \bibinfo {pages} {209} (\bibinfo {year} {2003})}\BibitemShut
  {NoStop}%
\bibitem [{\citenamefont {Tarasenko}(2006{\natexlab{a}})}]{tarasenko2006spin}%
  \BibitemOpen
  \bibfield  {author} {\bibinfo {author} {\bibfnamefont {S.~A.}\ \bibnamefont
  {Tarasenko}},\ }\href {\doibase 10.1103/PhysRevB.73.115317} {\bibfield
  {journal} {\bibinfo  {journal} {Phys. Rev. B}\ }\textbf {\bibinfo {volume}
  {73}},\ \bibinfo {pages} {115317} (\bibinfo {year}
  {2006}{\natexlab{a}})}\BibitemShut {NoStop}%
\bibitem [{\citenamefont
  {Tarasenko}(2006{\natexlab{b}})}]{tarasenko2006scattering}%
  \BibitemOpen
  \bibfield  {author} {\bibinfo {author} {\bibfnamefont {S.~A.}\ \bibnamefont
  {Tarasenko}},\ }\href {\doibase 10.1134/S0021364006160077} {\bibfield
  {journal} {\bibinfo  {journal} {JETP Letters}\ }\textbf {\bibinfo {volume}
  {84}},\ \bibinfo {pages} {199} (\bibinfo {year}
  {2006}{\natexlab{b}})}\BibitemShut {NoStop}%
\bibitem [{\citenamefont {Shen}\ \emph
  {et~al.}(2014{\natexlab{a}})\citenamefont {Shen}, \citenamefont {Raimondi},\
  and\ \citenamefont {Vignale}}]{shen2014theory}%
  \BibitemOpen
  \bibfield  {author} {\bibinfo {author} {\bibfnamefont {K.}~\bibnamefont
  {Shen}}, \bibinfo {author} {\bibfnamefont {R.}~\bibnamefont {Raimondi}}, \
  and\ \bibinfo {author} {\bibfnamefont {G.}~\bibnamefont {Vignale}},\ }\href
  {\doibase 10.1103/PhysRevB.90.245302} {\bibfield  {journal} {\bibinfo
  {journal} {Phys. Rev. B}\ }\textbf {\bibinfo {volume} {90}},\ \bibinfo
  {pages} {245302} (\bibinfo {year} {2014}{\natexlab{a}})}\BibitemShut
  {NoStop}%
\bibitem [{\citenamefont {Raimondi}\ \emph {et~al.}(2012)\citenamefont
  {Raimondi}, \citenamefont {Schwab}, \citenamefont {Gorini},\ and\
  \citenamefont {Vignale}}]{raimondi2012su2}%
  \BibitemOpen
  \bibfield  {author} {\bibinfo {author} {\bibfnamefont {R.}~\bibnamefont
  {Raimondi}}, \bibinfo {author} {\bibfnamefont {P.}~\bibnamefont {Schwab}},
  \bibinfo {author} {\bibfnamefont {C.}~\bibnamefont {Gorini}}, \ and\ \bibinfo
  {author} {\bibfnamefont {G.}~\bibnamefont {Vignale}},\ }\href {\doibase
  10.1002/andp.201100253} {\bibfield  {journal} {\bibinfo  {journal} {Annalen
  der Physik}\ }\textbf {\bibinfo {volume} {524}},\ \bibinfo {pages} {n/a}
  (\bibinfo {year} {2012})}\BibitemShut {NoStop}%
\bibitem [{\citenamefont {Raimondi}\ \emph {et~al.}(2006)\citenamefont
  {Raimondi}, \citenamefont {Gorini}, \citenamefont {Schwab},\ and\
  \citenamefont {Dzierzawa}}]{PhysRevB.74.035340}%
  \BibitemOpen
  \bibfield  {author} {\bibinfo {author} {\bibfnamefont {R.}~\bibnamefont
  {Raimondi}}, \bibinfo {author} {\bibfnamefont {C.}~\bibnamefont {Gorini}},
  \bibinfo {author} {\bibfnamefont {P.}~\bibnamefont {Schwab}}, \ and\ \bibinfo
  {author} {\bibfnamefont {M.}~\bibnamefont {Dzierzawa}},\ }\href {\doibase
  10.1103/PhysRevB.74.035340} {\bibfield  {journal} {\bibinfo  {journal} {Phys.
  Rev. B}\ }\textbf {\bibinfo {volume} {74}},\ \bibinfo {pages} {035340}
  (\bibinfo {year} {2006})}\BibitemShut {NoStop}%
\bibitem [{\citenamefont {Bindel}\ \emph {et~al.}(2016)\citenamefont {Bindel},
  \citenamefont {Pezzotta}, \citenamefont {Ulrich}, \citenamefont {Liebmann},
  \citenamefont {Sherman},\ and\ \citenamefont
  {Morgenstern}}]{bindel2016probing}%
  \BibitemOpen
  \bibfield  {author} {\bibinfo {author} {\bibfnamefont {J.~R.}\ \bibnamefont
  {Bindel}}, \bibinfo {author} {\bibfnamefont {M.}~\bibnamefont {Pezzotta}},
  \bibinfo {author} {\bibfnamefont {J.}~\bibnamefont {Ulrich}}, \bibinfo
  {author} {\bibfnamefont {M.}~\bibnamefont {Liebmann}}, \bibinfo {author}
  {\bibfnamefont {E.~Y.}\ \bibnamefont {Sherman}}, \ and\ \bibinfo {author}
  {\bibfnamefont {M.}~\bibnamefont {Morgenstern}},\ }\href@noop {} {\bibfield
  {journal} {\bibinfo  {journal} {Nature Physics}\ }\textbf {\bibinfo {volume}
  {12}},\ \bibinfo {pages} {920} (\bibinfo {year} {2016})}\BibitemShut
  {NoStop}%
\bibitem [{\citenamefont {Luttinger}\ and\ \citenamefont
  {Kohn}(1958)}]{KohnLuttingerBTE}%
  \BibitemOpen
  \bibfield  {author} {\bibinfo {author} {\bibfnamefont {J.~M.}\ \bibnamefont
  {Luttinger}}\ and\ \bibinfo {author} {\bibfnamefont {W.}~\bibnamefont
  {Kohn}},\ }\href {\doibase 10.1103/PhysRev.109.1892} {\bibfield  {journal}
  {\bibinfo  {journal} {Phys. Rev.}\ }\textbf {\bibinfo {volume} {109}},\
  \bibinfo {pages} {1892} (\bibinfo {year} {1958})}\BibitemShut {NoStop}%
\bibitem [{\citenamefont {Huang}\ \emph
  {et~al.}(2016{\natexlab{b}})\citenamefont {Huang}, \citenamefont {Chong},
  \citenamefont {Vignale},\ and\ \citenamefont
  {Cazalilla}}]{chunli2016graphene}%
  \BibitemOpen
  \bibfield  {author} {\bibinfo {author} {\bibfnamefont {C.}~\bibnamefont
  {Huang}}, \bibinfo {author} {\bibfnamefont {Y.~D.}\ \bibnamefont {Chong}},
  \bibinfo {author} {\bibfnamefont {G.}~\bibnamefont {Vignale}}, \ and\
  \bibinfo {author} {\bibfnamefont {M.~A.}\ \bibnamefont {Cazalilla}},\ }\href
  {\doibase 10.1103/PhysRevB.93.165429} {\bibfield  {journal} {\bibinfo
  {journal} {Phys. Rev. B}\ }\textbf {\bibinfo {volume} {93}},\ \bibinfo
  {pages} {165429} (\bibinfo {year} {2016}{\natexlab{b}})}\BibitemShut
  {NoStop}%
\bibitem [{\citenamefont {Landau}\ and\ \citenamefont
  {Lifshitz}(1980)}]{landau_lifshitz_sm1}%
  \BibitemOpen
  \bibfield  {author} {\bibinfo {author} {\bibfnamefont {L.~D.}\ \bibnamefont
  {Landau}}\ and\ \bibinfo {author} {\bibfnamefont {E.~M.}\ \bibnamefont
  {Lifshitz}},\ }\href@noop {} {\emph {\bibinfo {title} {Statistical Physics,
  Part I, Volume 5 of Course of Theoretical Physics}}}\ (\bibinfo  {publisher}
  {Pergamon},\ \bibinfo {year} {1980})\BibitemShut {NoStop}%
\bibitem [{\citenamefont {Shen}\ \emph
  {et~al.}(2014{\natexlab{b}})\citenamefont {Shen}, \citenamefont {Vignale},\
  and\ \citenamefont {Raimondi}}]{Kashen2014micro}%
  \BibitemOpen
  \bibfield  {author} {\bibinfo {author} {\bibfnamefont {K.}~\bibnamefont
  {Shen}}, \bibinfo {author} {\bibfnamefont {G.}~\bibnamefont {Vignale}}, \
  and\ \bibinfo {author} {\bibfnamefont {R.}~\bibnamefont {Raimondi}},\ }\href
  {\doibase 10.1103/PhysRevLett.112.096601} {\bibfield  {journal} {\bibinfo
  {journal} {Phys. Rev. Lett.}\ }\textbf {\bibinfo {volume} {112}},\ \bibinfo
  {pages} {096601} (\bibinfo {year} {2014}{\natexlab{b}})}\BibitemShut
  {NoStop}%
\bibitem [{\citenamefont {Kato}\ \emph
  {et~al.}(2004{\natexlab{a}})\citenamefont {Kato}, \citenamefont {Myers},
  \citenamefont {Gossard},\ and\ \citenamefont
  {Awschalom}}]{kato2004observation}%
  \BibitemOpen
  \bibfield  {author} {\bibinfo {author} {\bibfnamefont {Y.}~\bibnamefont
  {Kato}}, \bibinfo {author} {\bibfnamefont {R.}~\bibnamefont {Myers}},
  \bibinfo {author} {\bibfnamefont {A.}~\bibnamefont {Gossard}}, \ and\
  \bibinfo {author} {\bibfnamefont {D.}~\bibnamefont {Awschalom}},\ }\href@noop
  {} {\bibfield  {journal} {\bibinfo  {journal} {science}\ }\textbf {\bibinfo
  {volume} {306}},\ \bibinfo {pages} {1910} (\bibinfo {year}
  {2004}{\natexlab{a}})}\BibitemShut {NoStop}%
\bibitem [{\citenamefont {Valenzuela}\ and\ \citenamefont
  {Tinkham}(2006)}]{valenzuela2006direct}%
  \BibitemOpen
  \bibfield  {author} {\bibinfo {author} {\bibfnamefont {S.}~\bibnamefont
  {Valenzuela}}\ and\ \bibinfo {author} {\bibfnamefont {M.}~\bibnamefont
  {Tinkham}},\ }\href@noop {} {\bibfield  {journal} {\bibinfo  {journal}
  {Nature}\ }\textbf {\bibinfo {volume} {442}},\ \bibinfo {pages} {176}
  (\bibinfo {year} {2006})}\BibitemShut {NoStop}%
\bibitem [{\citenamefont {Avsar}\ \emph {et~al.}(2015)\citenamefont {Avsar},
  \citenamefont {Lee}, \citenamefont {Koon},\ and\ \citenamefont
  {Özyilmaz}}]{ahmet2015enhanced}%
  \BibitemOpen
  \bibfield  {author} {\bibinfo {author} {\bibfnamefont {A.}~\bibnamefont
  {Avsar}}, \bibinfo {author} {\bibfnamefont {J.~H.}\ \bibnamefont {Lee}},
  \bibinfo {author} {\bibfnamefont {G.~K.~W.}\ \bibnamefont {Koon}}, \ and\
  \bibinfo {author} {\bibfnamefont {B.}~\bibnamefont {Özyilmaz}},\ }\href
  {http://stacks.iop.org/2053-1583/2/i=4/a=044009} {\bibfield  {journal}
  {\bibinfo  {journal} {2D Materials}\ }\textbf {\bibinfo {volume} {2}},\
  \bibinfo {pages} {044009} (\bibinfo {year} {2015})}\BibitemShut {NoStop}%
\bibitem [{\citenamefont {S{\'a}nchez}\ \emph {et~al.}(2013)\citenamefont
  {S{\'a}nchez}, \citenamefont {Vila}, \citenamefont {Desfonds}, \citenamefont
  {Gambarelli}, \citenamefont {Attan{\'e}}, \citenamefont {De~Teresa},
  \citenamefont {Mag{\'e}n},\ and\ \citenamefont {Fert}}]{sanchez2013spin}%
  \BibitemOpen
  \bibfield  {author} {\bibinfo {author} {\bibfnamefont {J.~R.}\ \bibnamefont
  {S{\'a}nchez}}, \bibinfo {author} {\bibfnamefont {L.}~\bibnamefont {Vila}},
  \bibinfo {author} {\bibfnamefont {G.}~\bibnamefont {Desfonds}}, \bibinfo
  {author} {\bibfnamefont {S.}~\bibnamefont {Gambarelli}}, \bibinfo {author}
  {\bibfnamefont {J.}~\bibnamefont {Attan{\'e}}}, \bibinfo {author}
  {\bibfnamefont {J.}~\bibnamefont {De~Teresa}}, \bibinfo {author}
  {\bibfnamefont {C.}~\bibnamefont {Mag{\'e}n}}, \ and\ \bibinfo {author}
  {\bibfnamefont {A.}~\bibnamefont {Fert}},\ }\href@noop {} {\bibfield
  {journal} {\bibinfo  {journal} {Nature communications}\ }\textbf {\bibinfo
  {volume} {4}} (\bibinfo {year} {2013})}\BibitemShut {NoStop}%
\bibitem [{\citenamefont {Kato}\ \emph
  {et~al.}(2004{\natexlab{b}})\citenamefont {Kato}, \citenamefont {Myers},
  \citenamefont {Gossard},\ and\ \citenamefont {Awschalom}}]{kato2004current}%
  \BibitemOpen
  \bibfield  {author} {\bibinfo {author} {\bibfnamefont {Y.~K.}\ \bibnamefont
  {Kato}}, \bibinfo {author} {\bibfnamefont {R.~C.}\ \bibnamefont {Myers}},
  \bibinfo {author} {\bibfnamefont {A.~C.}\ \bibnamefont {Gossard}}, \ and\
  \bibinfo {author} {\bibfnamefont {D.~D.}\ \bibnamefont {Awschalom}},\ }\href
  {\doibase 10.1103/PhysRevLett.93.176601} {\bibfield  {journal} {\bibinfo
  {journal} {Phys. Rev. Lett.}\ }\textbf {\bibinfo {volume} {93}},\ \bibinfo
  {pages} {176601} (\bibinfo {year} {2004}{\natexlab{b}})}\BibitemShut
  {NoStop}%
\bibitem [{\citenamefont {Sih}\ \emph {et~al.}(2005)\citenamefont {Sih},
  \citenamefont {Myers}, \citenamefont {Kato}, \citenamefont {Lau},
  \citenamefont {Gossard},\ and\ \citenamefont {Awschalom}}]{sih2005spatial}%
  \BibitemOpen
  \bibfield  {author} {\bibinfo {author} {\bibfnamefont {V.}~\bibnamefont
  {Sih}}, \bibinfo {author} {\bibfnamefont {R.}~\bibnamefont {Myers}}, \bibinfo
  {author} {\bibfnamefont {Y.}~\bibnamefont {Kato}}, \bibinfo {author}
  {\bibfnamefont {W.}~\bibnamefont {Lau}}, \bibinfo {author} {\bibfnamefont
  {A.}~\bibnamefont {Gossard}}, \ and\ \bibinfo {author} {\bibfnamefont
  {D.}~\bibnamefont {Awschalom}},\ }\href@noop {} {\bibfield  {journal}
  {\bibinfo  {journal} {Nature Physics}\ }\textbf {\bibinfo {volume} {1}},\
  \bibinfo {pages} {31} (\bibinfo {year} {2005})}\BibitemShut {NoStop}%
\bibitem [{\citenamefont {Abanin}\ \emph {et~al.}(2009)\citenamefont {Abanin},
  \citenamefont {Shytov}, \citenamefont {Levitov},\ and\ \citenamefont
  {Halperin}}]{abanin2009nonlocal}%
  \BibitemOpen
  \bibfield  {author} {\bibinfo {author} {\bibfnamefont {D.~A.}\ \bibnamefont
  {Abanin}}, \bibinfo {author} {\bibfnamefont {A.~V.}\ \bibnamefont {Shytov}},
  \bibinfo {author} {\bibfnamefont {L.~S.}\ \bibnamefont {Levitov}}, \ and\
  \bibinfo {author} {\bibfnamefont {B.~I.}\ \bibnamefont {Halperin}},\ }\href
  {\doibase 10.1103/PhysRevB.79.035304} {\bibfield  {journal} {\bibinfo
  {journal} {Phys. Rev. B}\ }\textbf {\bibinfo {volume} {79}},\ \bibinfo
  {pages} {035304} (\bibinfo {year} {2009})}\BibitemShut {NoStop}%
\bibitem [{\citenamefont {Yang}\ \emph
  {et~al.}(2016{\natexlab{b}})\citenamefont {Yang}, \citenamefont {Huang},
  \citenamefont {Ochoa},\ and\ \citenamefont {Cazalilla}}]{hy2015extrinsic}%
  \BibitemOpen
  \bibfield  {author} {\bibinfo {author} {\bibfnamefont {H.-Y.}\ \bibnamefont
  {Yang}}, \bibinfo {author} {\bibfnamefont {C.}~\bibnamefont {Huang}},
  \bibinfo {author} {\bibfnamefont {H.}~\bibnamefont {Ochoa}}, \ and\ \bibinfo
  {author} {\bibfnamefont {M.~A.}\ \bibnamefont {Cazalilla}},\ }\href {\doibase
  10.1103/PhysRevB.93.085418} {\bibfield  {journal} {\bibinfo  {journal} {Phys.
  Rev. B}\ }\textbf {\bibinfo {volume} {93}},\ \bibinfo {pages} {085418}
  (\bibinfo {year} {2016}{\natexlab{b}})}\BibitemShut {NoStop}%
\bibitem [{\citenamefont {Han}\ \emph {et~al.}(2011)\citenamefont {Han},
  \citenamefont {Zhu}, \citenamefont {Zhang}, \citenamefont {Tan},
  \citenamefont {Ni},\ and\ \citenamefont {Niu}}]{han2011temperature}%
  \BibitemOpen
  \bibfield  {author} {\bibinfo {author} {\bibfnamefont {L.}~\bibnamefont
  {Han}}, \bibinfo {author} {\bibfnamefont {Y.}~\bibnamefont {Zhu}}, \bibinfo
  {author} {\bibfnamefont {X.}~\bibnamefont {Zhang}}, \bibinfo {author}
  {\bibfnamefont {P.}~\bibnamefont {Tan}}, \bibinfo {author} {\bibfnamefont
  {H.}~\bibnamefont {Ni}}, \ and\ \bibinfo {author} {\bibfnamefont
  {Z.}~\bibnamefont {Niu}},\ }\href@noop {} {\bibfield  {journal} {\bibinfo
  {journal} {Nanoscale research letters}\ }\textbf {\bibinfo {volume} {6}},\
  \bibinfo {pages} {84} (\bibinfo {year} {2011})}\BibitemShut {NoStop}%
\bibitem [{\citenamefont {Rammer}\ and\ \citenamefont
  {Smith}(1986)}]{rammer1986rmp}%
  \BibitemOpen
  \bibfield  {author} {\bibinfo {author} {\bibfnamefont {J.}~\bibnamefont
  {Rammer}}\ and\ \bibinfo {author} {\bibfnamefont {H.}~\bibnamefont {Smith}},\
  }\href {\doibase 10.1103/RevModPhys.58.323} {\bibfield  {journal} {\bibinfo
  {journal} {Rev. Mod. Phys.}\ }\textbf {\bibinfo {volume} {58}},\ \bibinfo
  {pages} {323} (\bibinfo {year} {1986})}\BibitemShut {NoStop}%
\bibitem [{\citenamefont {Rammer}(2011)}]{rammer_book}%
  \BibitemOpen
  \bibfield  {author} {\bibinfo {author} {\bibfnamefont {J.}~\bibnamefont
  {Rammer}},\ }\href@noop {} {\emph {\bibinfo {title} {Quantum Field Theory of
  Non-equilibrium States}}}\ (\bibinfo  {publisher} {Cambridge University
  Press},\ \bibinfo {year} {2011})\BibitemShut {NoStop}%
\end{thebibliography}%

\end{document}